\documentclass[a4paper,onecolumn,11pt,accepted=2020-09-17]{quantumarticle}
\pdfoutput=1
\usepackage[utf8]{inputenc}
\usepackage[english]{babel}
\usepackage[T1]{fontenc}
\usepackage{amsmath}

\usepackage{hyperref}
\usepackage[numbers]{natbib}

\usepackage{braket}
\usepackage{tikz}
\usepackage{lipsum}

\usepackage{array}
\usepackage{arydshln}
\usepackage{framed}
\usepackage{tcolorbox}
\begin{document}

\title{QUANTUM COMPUTING WITH NEUTRAL ATOMS}

\author{Lo\"{i}c Henriet}
\affiliation{Pasqal, 2 avenue Augustin Fresnel, 91120 Palaiseau, France}

\author{Lucas Beguin}
\affiliation{Pasqal, 2 avenue Augustin Fresnel, 91120 Palaiseau, France}

\author{Adrien Signoles}
\affiliation{Pasqal, 2 avenue Augustin Fresnel, 91120 Palaiseau, France}

\author{Thierry Lahaye}
\affiliation{Universit\'{e} Paris-Saclay, Institut d'Optique Graduate School, CNRS, Laboratoire Charles Fabry, 91127 Palaiseau Cedex, France}
\affiliation{Pasqal, 2 avenue Augustin Fresnel, 91120 Palaiseau, France}

\author{Antoine Browaeys}
\affiliation{Universit\'{e} Paris-Saclay, Institut d'Optique Graduate School, CNRS, Laboratoire Charles Fabry, 91127 Palaiseau Cedex, France}
\affiliation{Pasqal, 2 avenue Augustin Fresnel, 91120 Palaiseau, France}

\author{Georges-Olivier Reymond}
\affiliation{Pasqal, 2 avenue Augustin Fresnel, 91120 Palaiseau, France}

\author{Christophe Jurczak}
\affiliation{Pasqal, 2 avenue Augustin Fresnel, 91120 Palaiseau, France}
\affiliation{Quantonation, 58 rue d'Hauteville, 75010 Paris, France}

\maketitle

\begin{abstract}
The manipulation of neutral atoms by light is at the heart of countless scientific discoveries in the field of quantum physics in the last three decades. The level of control that has been achieved at the single particle level within arrays of optical traps, while preserving the fundamental properties of quantum matter (coherence, entanglement, superposition), makes these technologies prime candidates to implement disruptive computation paradigms. In this paper, we review the main characteristics of these devices from atoms / qubits to application interfaces, and propose a classification of a wide variety of tasks that can already be addressed in a computationally efficient manner in the Noisy Intermediate Scale Quantum\cite{Preskill_NISQ} era we are in. We illustrate how applications ranging from optimization challenges to simulation of quantum systems can be explored either at the digital level (programming gate-based circuits) or at the analog level (programming Hamiltonian sequences). We give evidence of the intrinsic scalability of neutral atom quantum processors in the 100-1,000 qubits range and introduce prospects for universal fault tolerant quantum computing and applications beyond quantum computing.
\end{abstract}



\section{Introduction}

One promising solution to the shortcomings of classical computing systems consists in the utilization of specialized accelerators for dedicated purposes. Examples of this type of computing systems include Graphics Processing Units (GPU), originally designed to accelerate graphics tasks like image rendering, or Field-Programmable Gate Arrays (FPGAs). Both technologies now provide high performance pipelines for efficient and high-speed computations, and are widely used as application-specific co-processors. Other types of processors are emerging with a special focus on acceleration of machine learning tasks. \\

In this respect, quantum co-processors, which manipulate the information at the quantum level, hold great promise for the future. Unlike classical information carried by digital bits 0 or 1, quantum information is commonly encoded onto a collection of two-level quantum systems referred to as qubits. A register containing $n$ qubits is described by a large complex vector of dimension 2$^n$. By taking advantage of the exponential amount of information available in quantum registers through superposition and entanglement, several quantum algorithms have been proposed that could outperform state-of-the-art classical algorithms for specific computing tasks. The most prominent example is Shor's factoring algorithm. While factoring a large integer $N$ into the product of its prime factors is a very hard task on a classical computer, it can in theory be done quite efficiently on a quantum device. More specifically, Shor's algorithm runs in a time increasing as a polynomial of $\log N$, whereas the best classical algorithm for this tasks requires sub-exponential time. \\

In such quantum algorithms, computation is described as a sequential application of quantum logic gates onto the qubits. This so-called digital approach to quantum information processing presents several key advantages, such as universality, as any operation can always be re-written with a finite subset of basis gates, and cross-device compatibility. Although the actual implementation of these quantum algorithms is still years away, as they would require ideal digital quantum processors, active research is currently exploring the capabilities of currently available quantum devices, which are imperfect and comprise a relatively modest number of qubits. These devices have recently shown that their computing capabilities can outperform classical supercomputers for a specific computing task\,\cite{Arute2019}. This opens up the so-called Noisy Intermediate Scale Quantum (NISQ) computing era\,\cite{Preskill_NISQ}, where 50-1,000 qubits devices allow for the exploration of the entanglement frontier, beyond what classical computers can do. \\

In order to build a viable quantum processor, a broad variety of physical platforms are currently being investigated\,\cite{Ladd10}. Among them, arrays of single neutral atoms manipulated by light beams appear as a very powerful and scalable technology to manipulate quantum registers with up to a few thousands of qubits\,\cite{saffman_quantum_2010,Saffman16}. In such quantum processors, each qubit is encoded into two electronic states of an atom. Nature provides that all the qubits are strictly identical when taken independently, as opposed to artifical atoms such as superconducting circuits or Silicon spin qubits that need to be manufactured with as little heterogeneity as possible. This feature is a remarkable advantage for neutral atom quantum processors to achieve low error rates during the computation. Atomic devices present other clear advantages with respect to other platforms, such as a large connectivity and the ability to natively realize multi-qubit gates.\\

In addition to the digital mode where the time evolution of qubits is described by quantum gates, control over the device can be realized at the so-called analog level, where the user can directly manipulate the mathematical operator (the Hamiltonian) describing the evolution of the ensemble of atoms. Not only does it allow for a finer level of control over pulses during the application of gates, but it also makes it possible to directly use the Hamiltonian of the system as a resource for computation. The fine level of control allowed by this analog setting, together with the large number of possible configurations, makes it a powerful  tool for quantum processing.\\ 

\begin{figure}[h!]
\center
\includegraphics[width=0.9\textwidth]{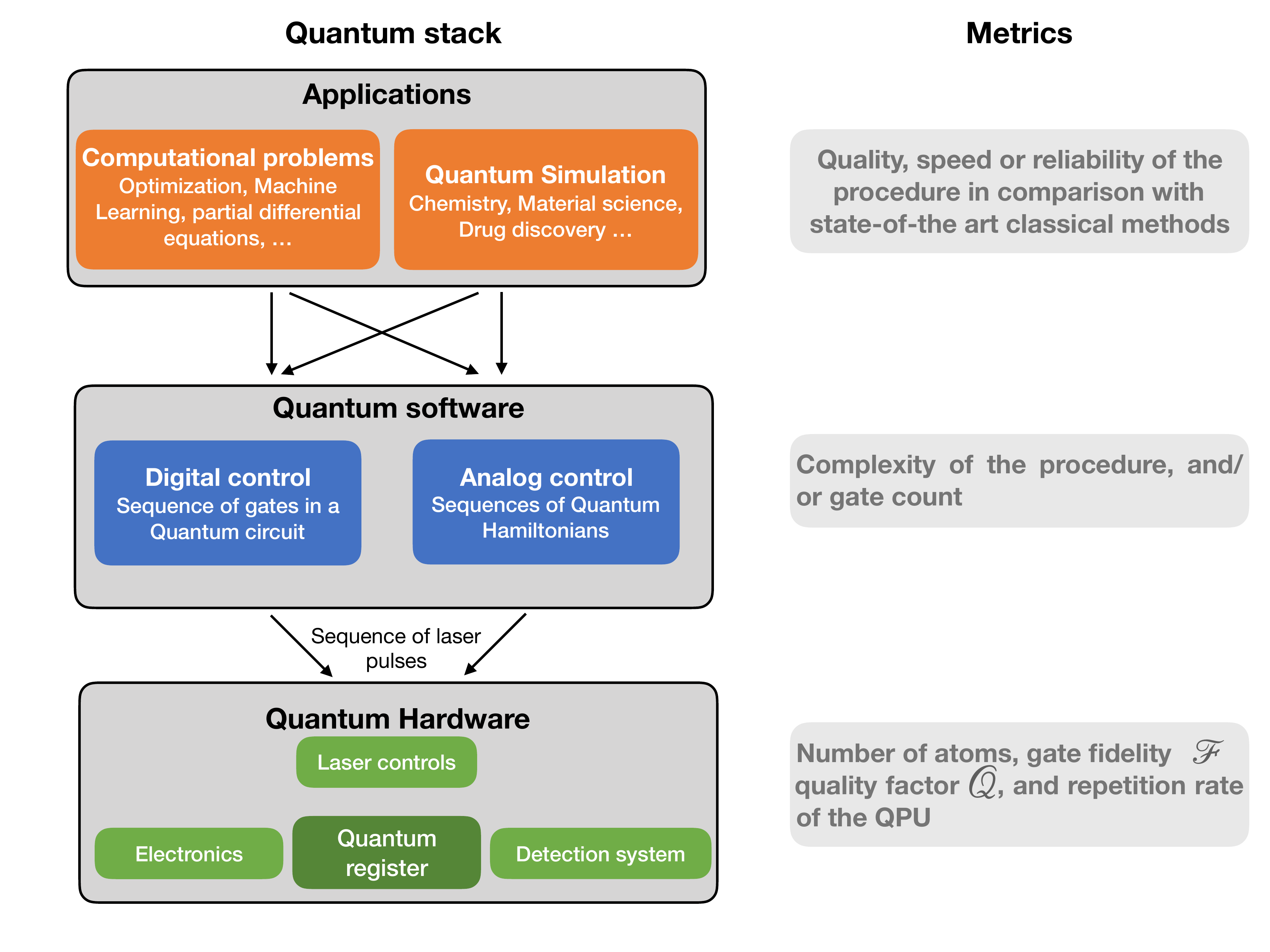}
\caption{Quantum stack and metrics to evaluate the performances of the quantum processor. Applications (top block of the Quantum stack) can be divided into two groups: Quantum Simulation problems that involve the study of a quantum system, and standard computational problems. To solve these problems, a quantum processor can either be used in a digital way or in an analog way (middle block of the Quantum stack). The low-level part of the stack corresponds to the physical quantum processor. 
}
\label{fig:intro-stack}
\end{figure}

Fundamental research on quantum information processing platforms using neutral atoms has been going on for years, and has led to impressive scientific results such as the simulation of highly complex quantum systems well above 100 qubits, beyond the reach of classical high performance computers\,\cite{browaeys2020many}. Only recently did it become possible to contemplate manufacturing devices for commercial use thanks to continuous progress in design and engineering. 
This technology will be described in more detail in Section \ref{sec:arrays}, with a particular emphasis on the technological choices taken by the team of A. Browaeys and T. Lahaye in an academic setting and at the company Pasqal.\\

Such a powerful machine can already bring value to several different fields, which will be described in Section\,\ref{sec:applications} (see top block of Fig.\,\ref{fig:intro-stack}). The most promising application is Quantum Simulation, where the quantum processor is used to gain knowledge over a quantum system of interest. As Richard Feynman already pointed out in the last century, it seems natural to use a quantum system as a computational resource for quantum problems. Pure science discovery will benefit from neutral atom quantum processors, and fields of applications are numerous at the industrial level, including for example the engineering of new materials for energy storage and transport, or chemistry calculations for drug discovery. Beyond quantum simulation, neutral atom quantum devices can also be used to solve hard computational problems, that cannot be efficiently solved on classical devices. Applications of NISQ computing notably encompass finding approximate solutions to hard combinatorial optimization problems\,\cite{Farhi14}, or enhancing the performances of Machine Learning procedures\,\cite{Schuld2018SupervisedLW}.  \\



The multiple technologies that will be developed will in the longer term be leveraged for the development of fault tolerant quantum computers, but also quantum networking and metrology. These perspectives will be briefly touched upon in Section~\ref{sec:Universal}.

\section{Neutral atoms arrays\label{sec:arrays}}
In this Section, we review the key technological blocks of neutral atom quantum processors, based on configurable arrays of single neutral atoms. The array can be seen as a register, where each single atom plays the role of a qubit. One common choice is rubidium atoms, a very common species in atomic physics that benefits from well-established technological solutions, especially in terms of lasers. More precisely, two electronic levels of the rubidium atoms are chosen to be the two qubit states, which we refer to as $\ket{0}$ and $\ket{1}$. Since the number of electronic states in an atom is infinite, there are various possible choices for implementing the qubit, leading to a very rich variety of configurations.\\

\begin{figure}[h!]
\center
\includegraphics[width=0.9\textwidth]{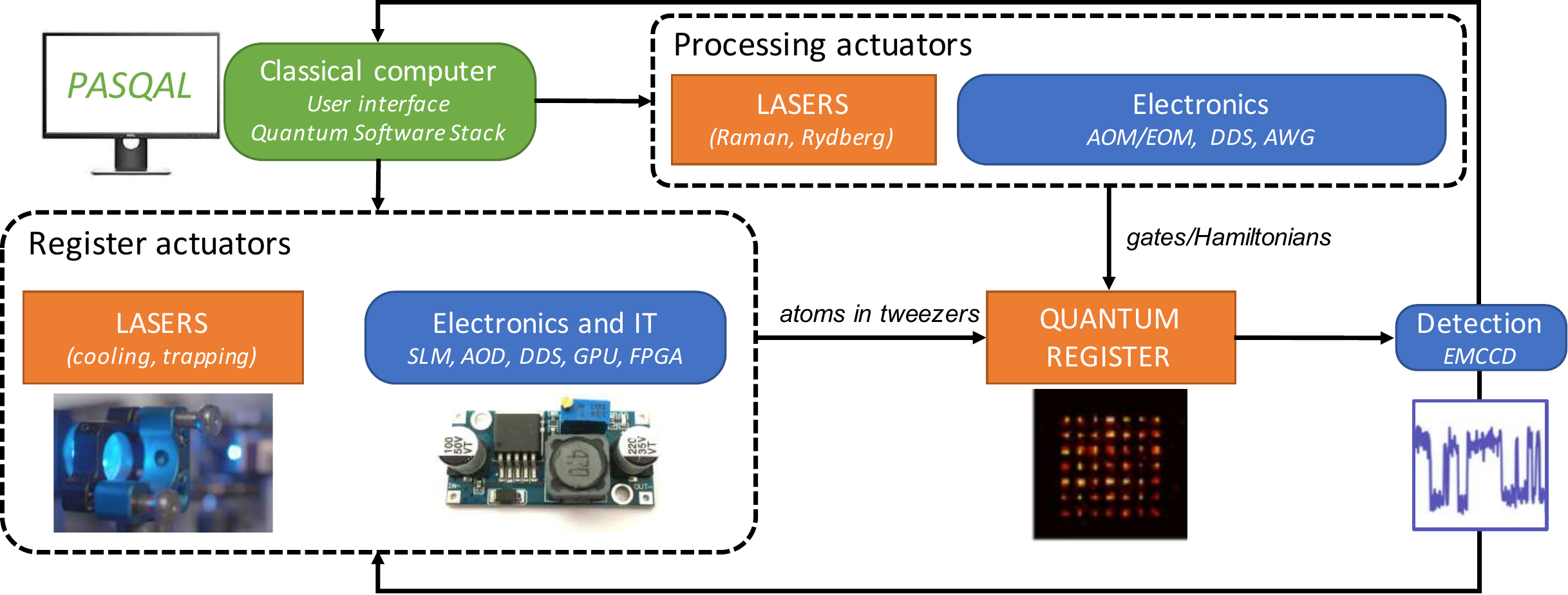}
\caption{Schematic of the hardware components of a neutral atom quantum device. The user sends, through the quantum software stack, instructions to the register actuators, which initialize the quantum register, and to the processing actuators, which perform the computation. Information in the quantum register is extracted through detection of an image. It serves as an input for real-time rearranging of the register and as an output of the computation.}
\label{fig:blocks}
\end{figure}

Light is the main tool to control both the position and the quantum state of the atoms. It is used to:
\begin{itemize}
    \item assemble and read-out registers made of hundreds of qubits (see Section~\ref{sec:register}),
    \item perform fully programmable quantum processing (see Section~\ref{sec:processing}).
\end{itemize}
For each dedicated task, a laser with a specific wavelength is required. In addition, electronic controls are needed to tune the light properties, apply instructions arising from the quantum software stack and extract information through atomic detection, as illustrated in Fig.~\ref{fig:blocks}. 

\subsection{Operating an atomic qubit register \label{sec:register}}

Unlike in solid quantum processors, e.g., superconducting, silicon or NV qubits, the register in an atomic QPU is not permanently built but is reconstructed after each processing. Hence, a typical computation cycle consists in three phases: register preparation, quantum processing and register readout. Although the processing stage is the one that receives most of the attention because it is where entanglement plays a role, the two other stages are of fundamental interest as well for developing a quantum processor. Indeed, highly efficient preparation and readout of the register have a direct impact on the performance of a quantum computing device.

\subsubsection{Register loading}

\begin{figure}[t!]
\center
\includegraphics[width=\textwidth]{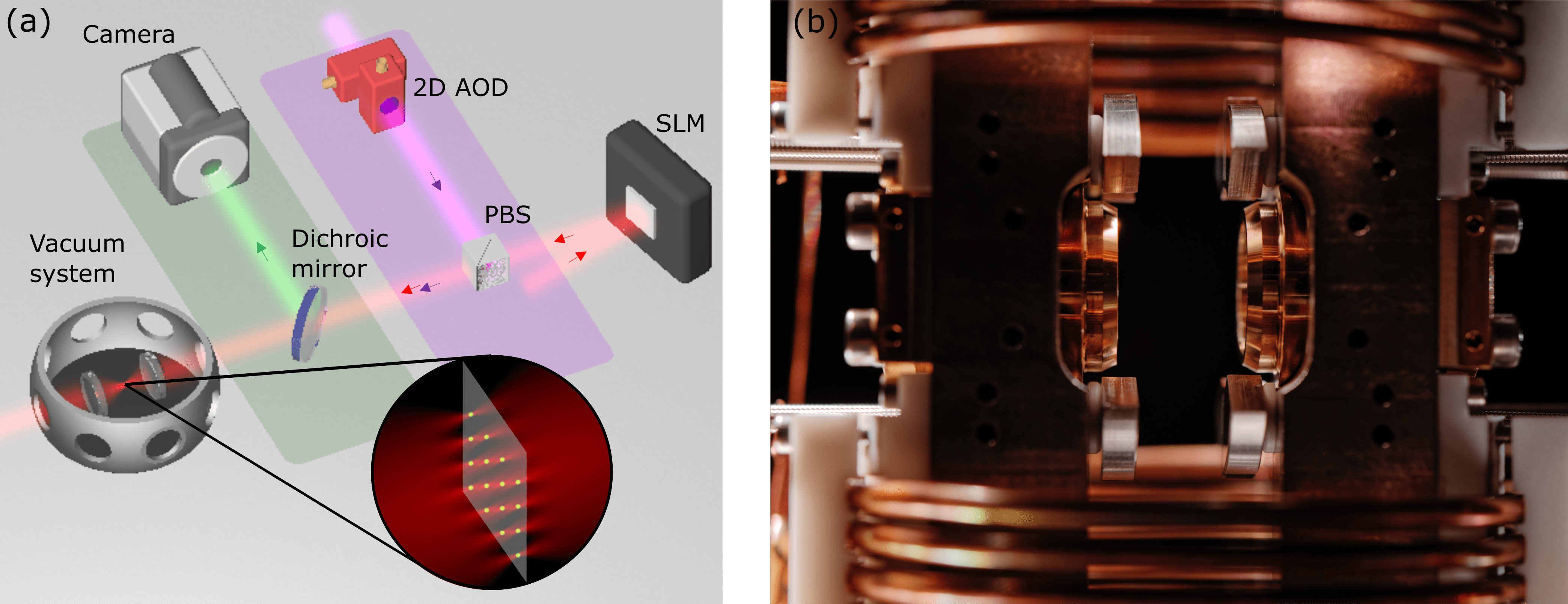}
\caption{(a) Overview of the main hardware components constituting a quantum processor. The trapping laser light (in red) is shaped by the spatial light modulator (SLM) to produce multiple microtraps at the focal plane of the lens (see inset). The moving tweezers (in purple), dedicated to rearranging the atoms in the register, are controlled by a 2D acousto-optic laser beam deflector (AOD) and superimposed on the main trapping beam with a polarizing beam-splitter (PBS). The fluorescence light (in green) emitted by the atoms is split from the trapping laser light by a dichroic mirror and collected onto a camera. (b) Photography of the heart of a neutral-atom quantum co-processor. The register is prepared at the center of this setup.}
\label{fig:setup}
\end{figure}

To prepare a register made of neutral atoms, one can use arrays of optical tweezers. It uses the hardware components shown in Fig.\ref{fig:setup}(a). 
As a starting point, a dilute atomic vapor is formed inside an ultra-high vacuum system operated at room temperature. With a first laser system (not shown), a cold ensemble of about $10^6$ atoms and 1 mm$^3$ volume is prepared inside a 3D magneto-optical trap (3D MOT), leveraging numerous laser cooling and trapping techniques~\cite{metcalf2003laser}. Then, a second trapping laser system isolates individual atoms within this ensemble. Using high numerical aperture lenses, the trapping beam gets strongly focused down to multiple spots of about 1$\,\mu$m diameter: the so-called optical tweezers \cite{schlosser2001sub}. Since the spots are only 10\,mm away from the lenses, the latter are placed inside the vacuum chamber (see Fig.~\ref{fig:setup}(b)). Within a trapping volume of a few $\mu$m$^3$, each tweezer contains at most one single atom at a time.\\

\begin{figure}[h!]
\center
\includegraphics[width=0.65\textwidth]{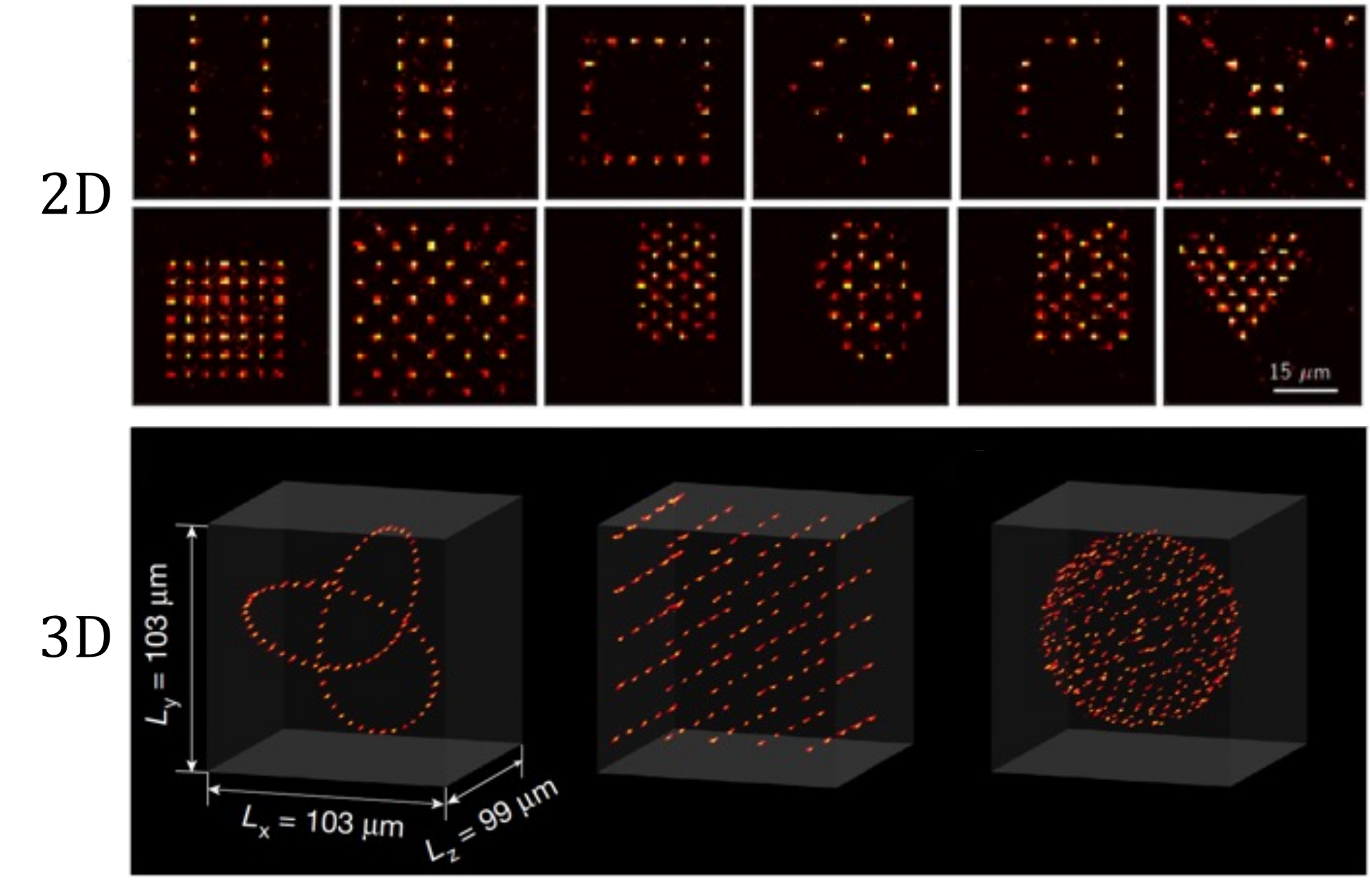}
\caption{Examples of tweezer arrays in 2D or 3D geometries, extracted from Ref~\cite{barredo_synthetic_2018}.}
\label{fig:arrays}
\end{figure}

The number of tweezers and their arrangement in any arbitrary 1D, 2D or even 3D geometries is fully tailored by holographic methods~\cite{nogrette2014single}. Before passing through the lens, the trapping beam is reflected onto a spatial light modulator (SLM) that imprints an adjustable phase pattern on the light. In the focal plane of the lens, the phase modulation gets converted into a desired intensity pattern, as illustrated in Fig.~\ref{fig:arrays}. For this reason, neutral atom platforms for quantum processing have a unique potential for scalability: the size of the quantum register is only limited by the amount of trapping laser power and by the performance of the optical system generating the optical tweezers.\\

Although each tweezer can host at most one atom, it happens only in about 50\% of the cases, the tweezer being empty otherwise. To detect which of the tweezers are filled, the atoms are imaged by collecting their fluorescence onto a sensitive camera (see Fig.\ref{fig:setup}(a)). The atoms are then moved from site to site in order to generate a pre-defined sub-register with unit filling. This operation is done using programmable moving optical tweezers, with a success rate above $99\%$. Elementary rearrangement steps are described in Fig.\ref{fig:rearrangement}(a). From the analysis of the initial image, an algorithm computes on the fly a set of moves to rearrange the initial configuration into the desired fully assembled sub-register. To implement this active feedback within few tens of milliseconds, the data are transferred through FPGA and the algorithm is run on an external GPU. The moves are then communicated to the drivers of 2D acousto-optic deflectors (AOD) that control the pointing and intensity of the moving tweezers. After rearrangement is completed, an assembled image is acquired to confirm the new positions of the atoms in the sub-register.

\subsubsection{Register readout}

Once the register is fully assembled, quantum processing can start. The next section describes in detail how to realize the required quantum operations. The processing by itself is extremely rapid since it happens in less than 100 $\mu$s, while the overall sequence, including loading and readout, lasts approximately 200~ms.\\ 

\begin{figure}[h!]
\center
\includegraphics[width=\textwidth]{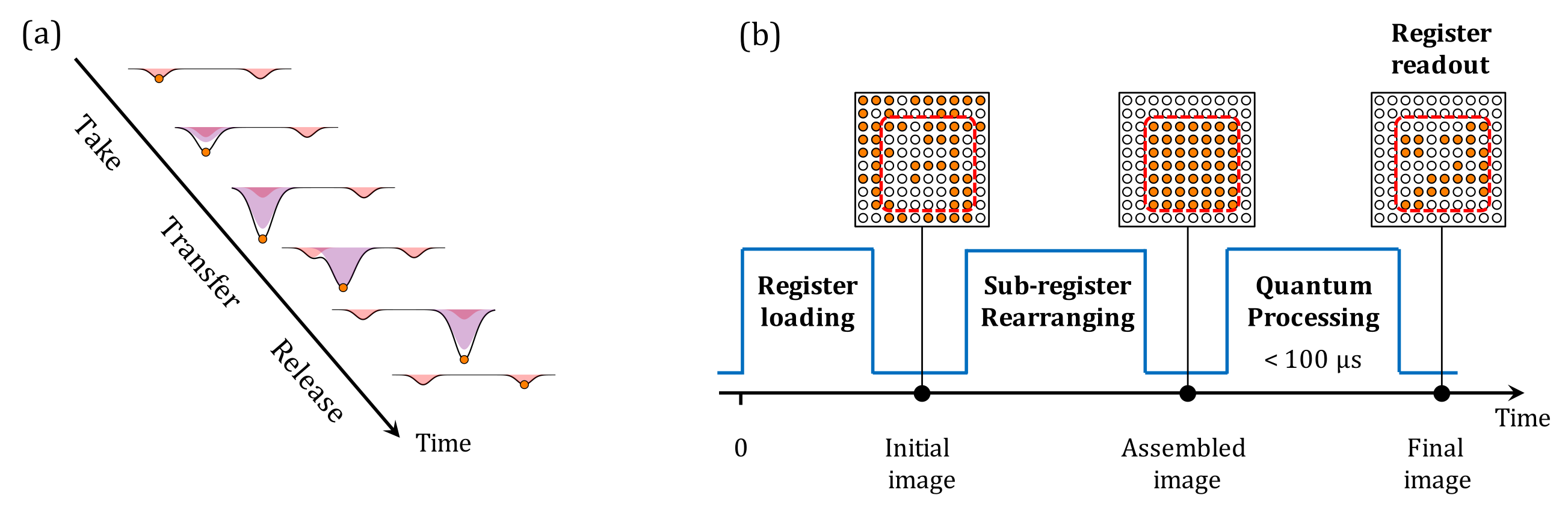}
\caption{(a) Moving a single atom from one site to another (both in red) in the register. The moving optical tweezer (in purple) first takes the atom, then transfers it and finally releases it into the other site. This operation takes less than 1 ms. (b) Temporal sequence of one computation cycle. The loading of the register being random, atoms are first rearranged to realize a defect-free sub-register, on which the quantum processing is performed.}
\label{fig:rearrangement}
\end{figure}

Once processing is done, the atomic register is read out by taking a final fluorescence image. It is performed such that each atom in qubit state $\ket{0}$ will appear as bright, whereas atoms in qubit state $\ket{1}$ remain dark, as illustrated in Fig.\ref{fig:rearrangement}(b). All the images are acquired on an electron-multiplying charge-coupled-device (EMCCD) camera: it converts with a very high sensitivity fluorescence photons into a bunch of electrons that produce a detectable electronic signal. Efficiencies of more than $98.6\%$ have been reported for this detection method~\cite{Fuhrmanek2011}, a value comparable to processors based on superconducting qubits~\cite{jeffrey2014fast}. Given the probabilistic nature of each possible outcome imposed by quantum mechanics, such computation cycles are then repeated many times in order to reconstruct the relevant statistical properties of the final quantum state produced by the algorithm.

\pagebreak
\subsection{Quantum processing with atomic qubits\label{sec:processing}}

\begin{figure}[h!]
\center
\includegraphics[width=\textwidth]{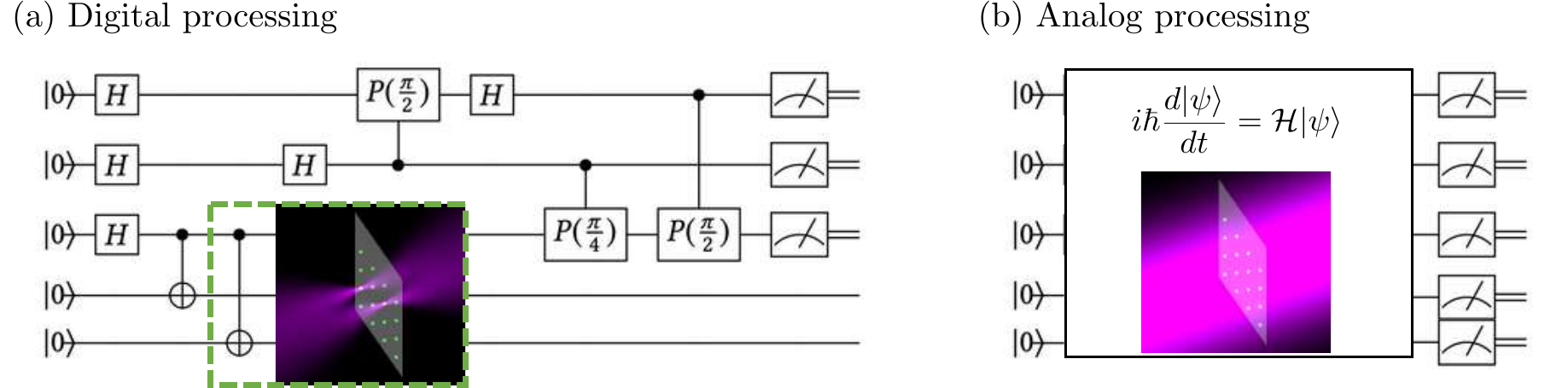}
\caption{Digital- vs analog processing. (a) In digital processing, a succession of gates is applied to the qubits to implement a quantum algorithm. Each gate is performed by addressing the qubits individually with laser beams. (b) In analog processing the qubits evolve under a tailored Hamiltonian $\mathcal{H}$, for instance by illuminating the whole register with a laser beam. The wavefunction $\ket{\psi}$ of the system follows the Schr\"{o}dinger equation.}
\label{fig:digital-analog}
\end{figure}

Neutral atom quantum processors (QPU) are able to implement both digital and analog quantum processing tasks. In digital computing, quantum algorithms are decomposed into a succession of quantum logic gates, described by a quantum circuit as shown in Fig.~\ref{fig:digital-analog}(a). The quantum gates are realized by shining fine-tuned laser pulses onto a chosen subset of individual atoms in the register. In analog computing, lasers are applied to realize an Hamiltonian. The qubits evolve in time according to the Schr\"{o}dinger equation, as illustrated in Fig.~\ref{fig:digital-analog}(b). The final state of the system is probed by measuring the state of each individual qubits. Implementation of digital and analog computing tasks are discussed in this section.

\subsubsection{Digital quantum processing\label{sec:gates}}
Digital computing requires qubits robust against decoherence, i.e., weakly coupled to their environment. Therefore one can use as a qubit the two hyperfine ground states ($F=1$ and $F=2$) of the rubidium atom, because they have very long lifetimes (of the order of tens of years) that prevent radiative coupling to the electromagnetic environment. As discussed earlier, gates are performed with laser beams. The spacing between atoms in the register being typically several micrometers, specific qubits can be addressed with high accuracy by strongly focusing the lasers (as illustrated in the inset of the CNOT gate in Fig.~\ref{fig:digital-analog}(a)). Interestingly, one- and two-qubit gates are all that is needed to constitute a universal gate set. For example, the ability to realize arbitrary single qubit rotations and the well-known two-qubit CNOT gate is sufficient for the realization of any quantum algorithm\,\cite{Nielsen}.\\

\begin{figure}[h!]
\center
\includegraphics[width=\textwidth]{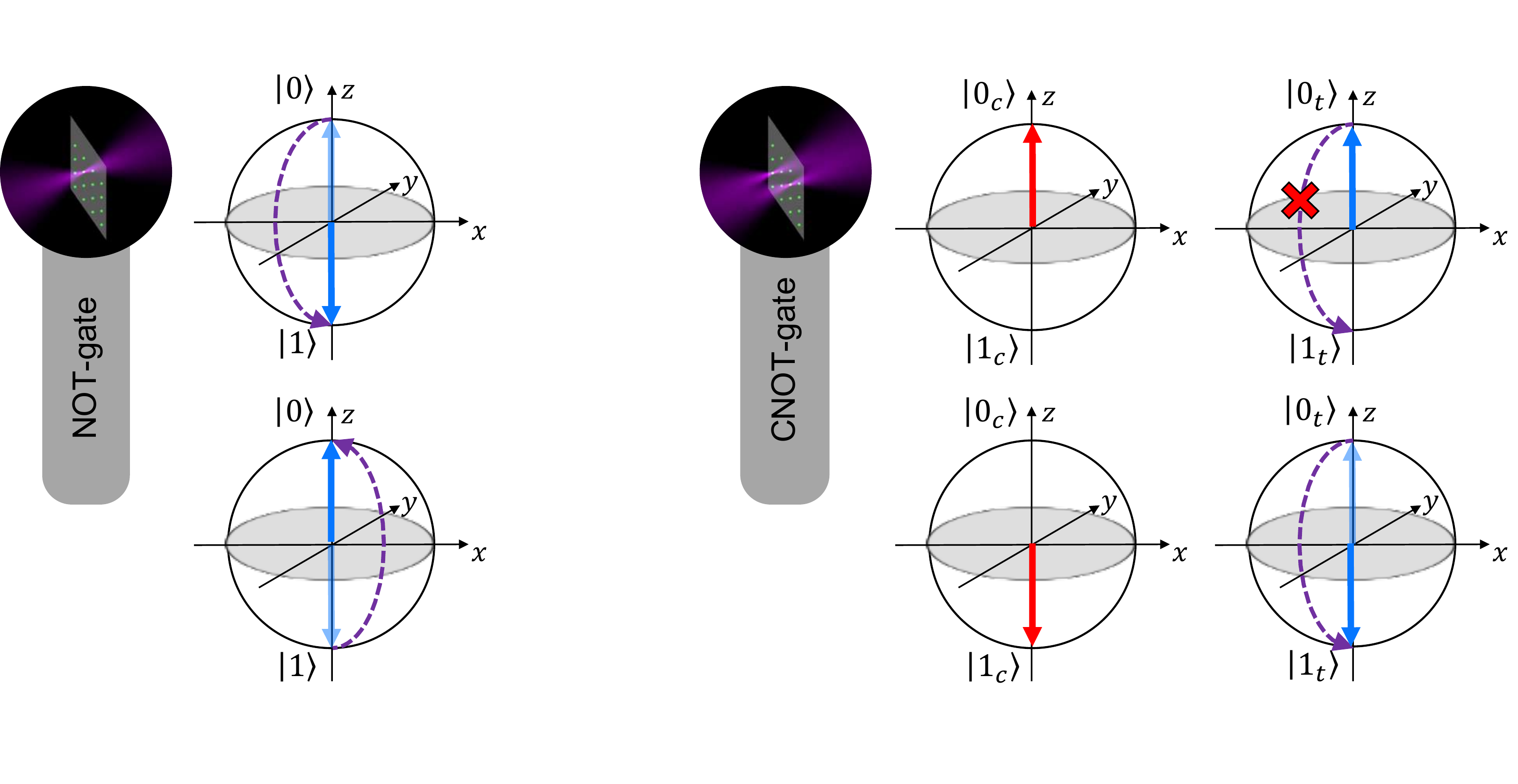}
\caption{Single- vs two-qubit gates. (a) Illustration of a NOT gate. When the addressed spin is in the state $\ket{0}$ (top sphere), the laser field rotates it by an angle of $\pi$ to end up in the state $\ket{1}$. When starting in $\ket{1}$, it ends up in $\ket{0}$ (bottom sphere). (b) According to the state of the control qubit (left column, in red), the target qubit (right column, in blue) is rotating by $\pi$ or not (illustrated in the case of a target qubit initially in $\ket{0}$.}
\label{fig:NOT_CNOT}
\end{figure}

One-qubit gates are specific unitary transformations described by 2-by-2 complex matrices transforming one qubit state into another. Notable examples are the NOT-gate that changes the state $\ket{0}$ into $\ket{1}$ and vice-versa, or the Hadamard (H) gate that generates superposition of both states starting from a pure state. In the $\{\ket{0},\ket{1}\}$ basis, theses gates read
\begin{align*}
\text{NOT}=\begin{bmatrix} 0 & 1 \\ 1 & 0 \end{bmatrix} \hspace{4cm} \text{H}= \frac{1}{\sqrt{2}}\begin{bmatrix} 1 & 1 \\ 1 & -1 \end{bmatrix}.
\end{align*}\\

A nice way to illustrate how a quantum gate modifies the state of a qubit is to use the Bloch sphere picture. In this picture, a qubit pure state is represented by a 3D unitary vector, the Bloch vector, pointing somewhere on the surface of a unit sphere, the Bloch sphere. The state $\ket{0}$ occupies the North pole and the state $\ket{1}$ the South pole, and any point at the surface of the sphere is associated with another unique qubit state, being a coherent superposition of $\ket{0}$ and $\ket{1}$. As unitary transformations, any single-qubit gate corresponds to a rotation of the Bloch vector onto the Bloch sphere. In this respect, they can always be written as combinations of the Pauli matrices $\sigma^x,\,\sigma^y$ and $\sigma^z$. Regarding the previous examples, the NOT-gate coincides with $\sigma^x$ and is a rotation by a $\pi$ angle around the $x$ axis (see Fig.~\ref{fig:NOT_CNOT}(a)), while the Hadamard-gate writes $(\sigma^x+\sigma^z)/\sqrt{2}$ and is a rotation of $\pi$ around the $(x+z)$ axis.\\

Arbitrary rotations of the qubit state on the Bloch sphere can be performed by driving the qubit transition with a control field. In our device the latter is an optical laser field driving Raman transitions through an intermediate atomic state~\cite{yavuz2006fast,Levine19}. The atom-laser interaction is characterized by the Rabi frequency $\Omega$ (its strength, proportional to the amplitude of the laser field), the detuning $\delta$ (the difference between the qubit resonance and the field frequencies) and their relative phase $\varphi$. Driving the control field for a duration $\tau$ induces rotations around the $(x,y,z)$ axes with angles $(\Omega\tau \cos{\varphi},\Omega\tau \sin{\varphi}, \delta\tau)$. Hence, any single-qubit gate can be implemented by tuning the pulse duration, the laser intensity, the detuning and the phase of the laser. These parameters are controlled using direct digital synthesizers (DDS) that drive acousto- and electro-optic modulators (AOM and EOM) placed on the laser beams. 
As an illustration, when driving the control field at resonance ($\delta=0$), the qubit oscillates in time between the states $\ket{0}$ and $\ket{1}$. These so-called Rabi oscillations are a general feature of quantum two-level systems, and therefore serve as a test for controlling the efficiency of single-qubit gates. In atomic registers they have been observed with an extremely high contrast~\cite{Levine19,madjarov2020high}, corresponding to single-qubit gate fidelities $\mathcal{F}$ higher than $99.5\%$. The fidelity $\mathcal{F}$ of a quantum operation is a number between 0 and 1 that measures the closeness between the state actually created by the operation and the theoretically expected state. A fidelity $\mathcal{F}=99.5\%$ corresponds to a probability of $0.5\%$ of measuring an undesired outcome after the operation. When stopped after half an oscillation, i.e. for a pulse area equal to $\pi$, it exactly realizes a NOT gate. On the other hand, the Hadamard gate requires a $\pi$ pulse at a detuning equal to the Rabi frequency ($\delta=\Omega$) and the laser phase equal to 0.\\


\definecolor{shadecolor}{rgb}{0.78,0.78,0.78}
\begin{shaded}
\noindent{\bf Box 1 $|$ Implementation of the CNOT-gate}
\end{shaded}
\vspace{-9mm}
\definecolor{shadecolor}{rgb}{0.835294,0.835294,0.835294}
\begin{shaded}

\noindent The CNOT gate flips a “target” qubit state if and only if a “control” qubit is in the state $\ket{1}$. The corresponding matrix in the pair state basis $\{\ket{0_c 0_t},\ket{0_c 1_t},\ket{1_c 0_t},\ket{1_c 1_t}\}$ is
\begin{align*}
\text{CNOT} = \begin{bmatrix} 1 & 0 & 0 & 0 \\ 0 & 1 & 0 & 0 \\ 0 & 0 & 0 & 1 \\ 0 & 0 & 1 & 0 \end{bmatrix}.
\end{align*}
There are several ways to realize this gate using dipolar Rydberg interaction~\cite{isenhower2010demonstration,Levine19}. Figure B1.1(a) illustrates the key mechanism with a simple sequence of 3 pulses.
\begin{itemize}
    \item With initial state $\ket{1_c 1_t}$, all the three pulses are off-resonant and the state remains unchanged.
    \item With initial state $\ket{0_c 1_t}$, pulse 2 is off-resonant so the target qubit state remains unchanged. Pulses 1 and 3 drive a 2$\pi$ rotation of the control qubit. The pair state picks up a phase factor $e^{i\pi}= -1$.
    \item With initial state $\ket{1_c 0_t}$, pulses 1 and 3 are off-resonant so the control qubit state remains unchanged. Pulse 2 drives a 2$\pi$ rotation of the target qubit. The pair state picks up a phase factor $e^{i\pi} = -1$.
    \item With initial state $\ket{0_c 0_t}$, pulse 1 excites the control qubit to state $\ket{r}$ with a $\pi$ rotation. Because of the Rydberg blockade, the target qubit Rydberg state $\ket{r}$ is shifted out of resonance during pulse 2. The target qubit remains in state $\ket{0_t}$. Finally, pulse 3 brings the control qubit back to state $\ket{0_c}$ with another $\pi$ rotation. The pair state picks up a factor $e^{i\pi}= -1$.
\end{itemize}
This pulse sequence realizes, within a global $\pi$ phase, the controlled-Z (CZ) gate
\begin{align*}
\begin{bmatrix} -1 & 0 & 0 & 0 \\ 0 & -1 & 0 & 0 \\ 0 & 0 & -1 & 0 \\ 0 & 0 & 0 & 1 \end{bmatrix} = e^{i\pi} \begin{bmatrix} 1 & 0 & 0 & 0 \\ 0 & 1 & 0 & 0 \\ 0 & 0 & 1 & 0 \\ 0 & 0 & 0 & -1 \end{bmatrix} = e^{i\pi} \text{CZ}.
\end{align*}
The CZ gate can then be used to generate the CNOT gate by including Hadamard gates on the target atom before and after the operation, as shown in Fig.B1.1(b).

\begin{center}
\includegraphics[width=0.9\textwidth]{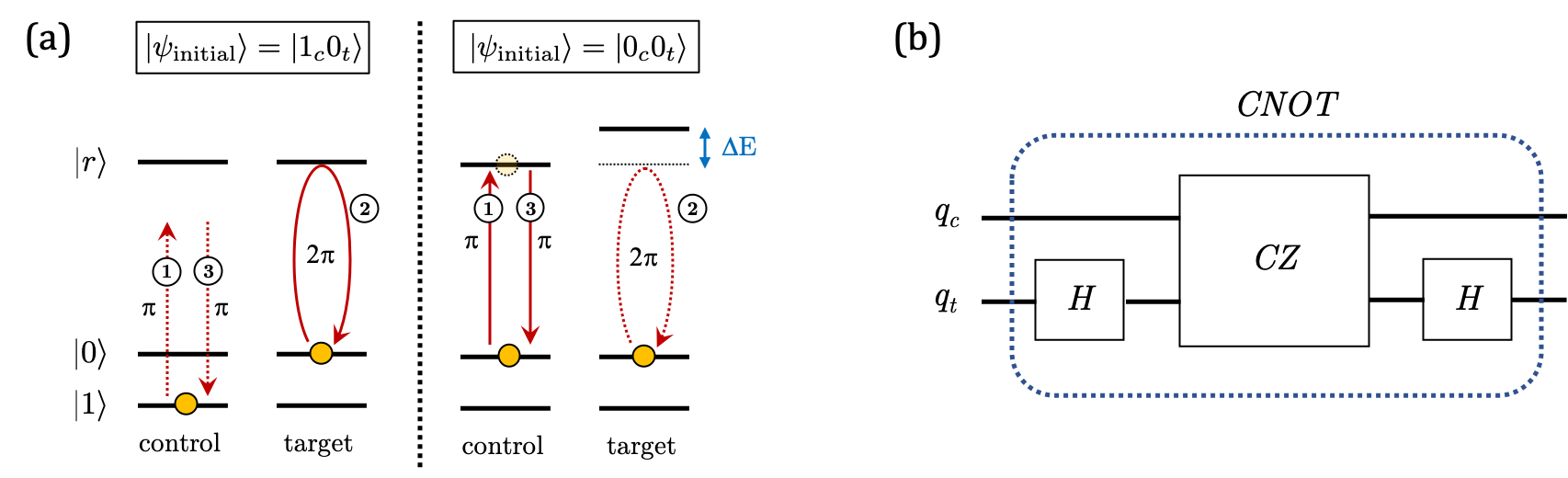}
\end{center}
\begin{small}
\noindent{\bf Figure B1.1 $|$ Implementation of the CNOT gate using Rydberg interactions.} 
(a) Principle of the controlled-Z gate based on dipolar Rydberg interaction. First a $\pi$ pulse is applied on the control atom, then a 2$\pi$ pulse on the target atom, and finally another $\pi$ pulse on the control one. (b) Realization of a CNOT gate using a CZ gate and two Hadamard gates. 
\end{small}

\end{shaded}

Two-qubit gates are unitary transformations described by 4-by-4 matrices that transform one two-qubit state into another. They are the most basic but crucial resource allowing the generation of entanglement in the register. Physically, their implementation requires an interaction between the qubits. However, neutral atoms in their electronic ground state can only interact significantly via contact collisions. Therefore, single atoms --typically separated by a few micrometers in the register-- do not naturally feel each other. In 2000, Jaksch et al.~\cite{jaksch_fast_2000} proposed a scheme that takes advantage of the dipole-dipole interaction between atoms when they are prepared to highly excited electronic states. In these so-called Rydberg states, the atoms exhibit a huge electric dipole moment that can be three orders of magnitude bigger than in their ground state. Therefore, two Rydberg atoms --separated by a few micrometers-- will experience a dipole-dipole interaction strong enough to shift significantly the energy of the doubly excited state, preventing the excitation of two atoms at the same time. This effect is called Rydberg blockade and is the basic mechanism to achieve a quantum logic: the excitation of a first atom to a Rydberg state conditions the excitation of a second one.\\

\begin{figure}[b!]
\center
\includegraphics[scale=0.28]{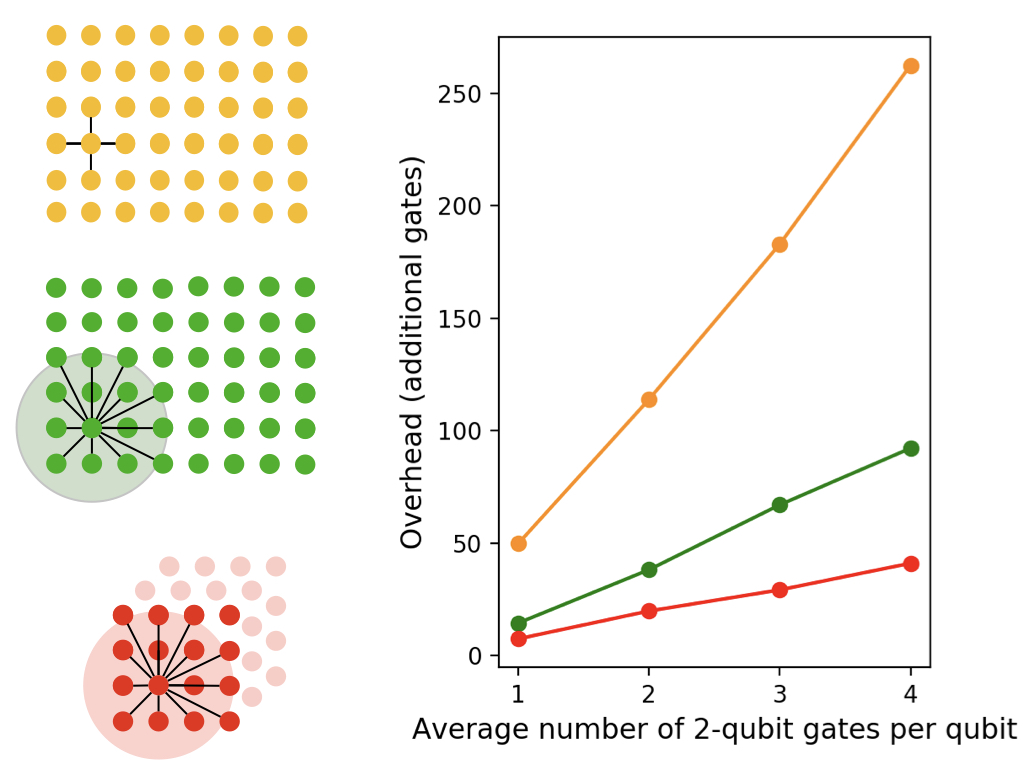}
\caption{Implementation of random quantum circuits on three distinct device architectures with 48 qubits, illustrated on the left. The orange layout displays nearest-neighbor connectivity in 2D, while the green layout has an interaction radius (Rydberg blockade radius) of 2.3 units of lattice spacing. The red layout corresponds to a 3D cubic lattice, with the same interaction radius. These characteristics result in a large difference in terms of the final number of gates necessary to realize quantum circuits. This is shown on the right panel, where we plot the average gate count overhead with respect to the number of two-qubit gates.  The data points were obtained using Google Cirq qubit routing compiler, and each point corresponds to an average over 64 random circuits.  }
\label{overhead-connectivity}
\end{figure}

Hence, Rydberg-mediated entanglement is particularly well suited to implement the controlled-NOT (CNOT) gate (see Fig.~\ref{fig:NOT_CNOT}(b)). Following the protocol described in Box 1, the gate can be performed within 1 $\mu$s, leading to a 1 MHz clock rate for the QPU. Transient non-zero population of Rydberg states may lead to errors in the process. Current imperfection levels in the experiment allow us to perform more than a hundred gates within the typical coherence time of the system. The experimentally achieved fidelities have been measured to be $\mathcal{F}=94.1\%$~\cite{levine_high-fidelity_2018} and will be largely improved. In particular, imperfections can be compensated by implementing optimal control~\cite{rach2015dressing}. The goal is to affect the dynamics in a controlled way in order to increase the probability of reaching the desired quantum state. It is performed by shaping the optical pulses in time instead of applying basic squared pulses. It calls specific algorithms that optimized the shapes to maximize the gate fidelity. Practically it requires extremely fast arbitrary waveform generators that are able to sample the pulses at the 100 ns timescale.\\

The ability to achieve two-qubit gates between distant qubits has critical consequences on the implementation of long quantum circuits. When a quantum processor suffers from a limited connectivity, gates that are required for the implementation of a given algorithm cannot be implemented natively. Circumventing this problem requires the introduction of SWAP gates\,\cite{Nielsen}, that exchange locally the quantum state between different qubits. This gate addition comes at a high cost, as each additional gate introduces errors along the computation. The relatively large connectivity of neutral atom devices allows to mitigate this effect.\\

To illustrate this strong advantage, we use Cirq built-in qubit routing compiler on a variety of random circuits to be implemented either on a device with nearest-neighbor connectivity such as most transmon architectures for superconducting qubits, or on a neutral atom device in 2D and 3D. Results are displayed on Fig.\,\ref{overhead-connectivity}, where we see that the overhead for neutral atom devices (green points for the 2D device and red points for the 3D device) is 5 to 10 times lower than the one obtained with a device with nearest-neighbor connectivity.\\

\begin{figure}[h!]
\center
\includegraphics[scale=0.45]{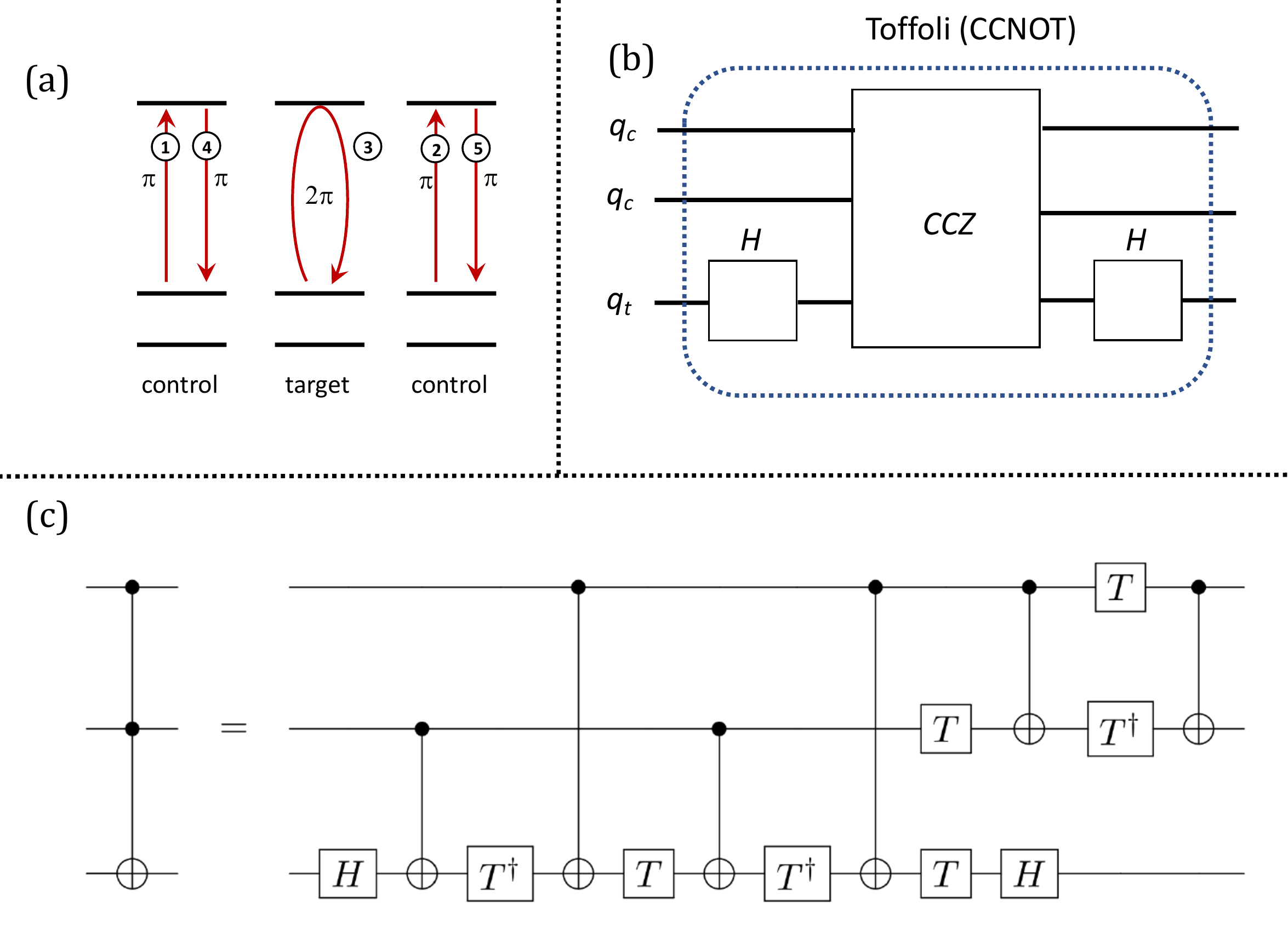}
\caption{(a) Sequence of 5 pulses realizing a CCZ gate. (b) Decomposition of the Toffoli gate (CCNOT) into a CCZ surrounded by two single-qubit gates. (c) The Toffoli gate can be decomposed into a minimum of six CNOTs, with additional single qubits gates (Hadamard gates H and T ($\pi/8$) gates). This last sequence if equivalent to a total of more than 30 laser pulses.}
\label{Toffoli}
\end{figure}

In addition to large connectivity, neutral atom devices present the advantage of natively implementing multi-qubit gates involving more that 2 qubits\,\cite{saffman_quantum_2010,Saffman16,Levine19,Khazali20}. Such multi-qubit gates are instrumental for the efficient implementation of several quantum algorithms, including for example Grover search or the variational resolution of non-linear partial differential equations\,\cite{Lubasch20}. In that respect, neutral atom quantum processors are for example able to realize a three-qubit Toffoli gate, which corresponds to a NOT operation applied on a target qubit conditioned on the states of two control qubits, at a modest cost in terms of operations to be performed by the processor. More specifically, a Toffoli gate can be achieved with a sequence of only 7 pulses (2 Hadamard gates on both sides of a 5-pulse CCZ, as illustrated on Fig.\,\ref{Toffoli} (a) and (b)). On platforms that cannot natively implement a Toffoli gate, one has to combine not less than six Controlled-NOT gates and 9 single-qubit gates to achieve the same effect, leading to a very large overhead as illustrated on Fig.\,\ref{Toffoli}. \\

These properties of large connectivity and native multi-qubit gates, specific to Rydberg-based gates, help reducing drastically the overall duration of the processing. It is of fundamental interest since, for the current generation of devices, the clock rate (relative to the computation, 1 MHz) and the acquisition rate (relative to the measurement, 5 Hz) in a neutral atom QPU are relatively small in comparison to other quantum computing platforms. Together with the 2D and 3D capabilities, all these features contribute to making neutral atom QPUs highly competitive. On top of that, what constitutes the main advantages of neutral atom QPUs is their very high scalability: having more qubits simply relies on the capability of increasing the number of tweezers in the interference pattern of the laser, but does not involve manufacturing new chips as it is the case with solid-state platforms where a key challenge is the reproducibility of each single qubit over a large device.

\subsubsection{Analog quantum processing \label{sec:AnalogProcessing}}

Neutral atoms arrays are also suitable to implement quantum Hamiltonians and realize analog quantum processing. Rydberg atoms behaving as huge electric dipoles, they experience dipole-dipole interactions that map into spin Hamiltonians. Each qubit of the register then behaves as a spin whose states are $\ket{\downarrow}=\ket{0}$ and $\ket{\uparrow}=\ket{1}$. Depending on the Rydberg states that are involved in the process, the spins experience different types of interactions that translate into different Hamiltonians~\cite{browaeys2020many}. The most studied one is the Ising model, which is obtained when the $\ket{\downarrow}$ state is one of the ground states and the $\ket{\uparrow}$ a Rydberg state~\cite{schauss2015crystallization,labuhn2016tunable,Bernien17,leseleuc2018accurate}. In this case the Hamiltonian writes
\begin{equation}
\mathcal{H}(t)= \frac{\hbar}{2} \Omega(t) \sum_j \sigma_j^x-\hbar \delta(t)\sum_j   n_j+\sum_{i \neq j}\frac{C_6}{r_{ij}^6} n_i n_j,
\label{eq:ising_hamiltonian}
\end{equation}
with $n_j=(1+\sigma_j^z)/2$ the Rydberg state occupancy and $\sigma^{x,z}_j$ the Pauli matrices $\sigma^{x,z}$ of the spin $j$. The first terms are induced by the laser that couples the qubit states and relate to an effective magnetic field, with transverse and longitudinal components $B_{\perp} \propto \Omega(t)$ and $B_{||} \propto -\delta(t)$. Physically, these two terms describe how individual magnets would precess in the presence of an external magnetic field. The parameters which control this precession are $\Omega(t)$ and $\delta(t)$, respectively the Rabi frequency and detuning. They can be varied by changing the intensity and frequency of the laser field. The third term in~\eqref{eq:ising_hamiltonian} relates to the interactions between individual spins. More specifically, it corresponds to an energy penalty that two qubits experience if they are both in the Rydberg states at the same time. The coupling between two spins $i$ and $j$ is of van der Waals type and depends on the inverse of the distance between them $r_{ij}$ to the power of 6, and on a coefficient $C_6$ relative to the Rydberg state (see Box 2).\\ 

The Ising Hamiltonian is a prototypical model that tackles many problems in condensed matter. For example, this is a model that describes how quantum magnets evolve at very low temperatures in material sciences. Interestingly, it is also relevant for a broad variety of classical applications as it can cast optimization problems that are computationally difficult to solve~\cite{lucas2014ising}, as further described in section~\ref{sec:applications}. In neutral atom devices, it can be implemented in 1D, 2D or 3D arrays containing few hundreds of atoms, far beyond the computational capabilities of classical computers.

\definecolor{shadecolor}{rgb}{0.78,0.78,0.78}
\begin{shaded}
\noindent{\bf Box 2 $|$ Dipolar vs van der Waals interactions}
\end{shaded}
\vspace{-9mm}
\definecolor{shadecolor}{rgb}{0.835294,0.835294,0.835294}
\begin{shaded}

\noindent Atoms in Rydberg states exhibit huge electric dipole moments. Two Rydberg atoms thus interact via the electric dipole-dipole interactions, whose Hamiltonian $V_{dd}$ is proportional to $1/R^3$, with $R$ the distance between the two atoms~\cite{Ravets2015}.

At the first order of perturbation theory, two pair states $\ket{\psi}$ and $\ket{\psi'}$ are coupled only if the matrix element $\bra{\psi}V_{dd}\ket{\psi^\prime}$ is not vanishing. It is the case when considering for instance the states $\ket{\psi}=\ket{nS,n'P}$ and $\ket{\psi^\prime}=\ket{n'P,nS}$, where {$n$, $n'$} are principal quantum numbers, and S and P represent states with orbital angular momentum $L$ respectively equals to 0 and 1. Since these two states have the same energy, a pair of atoms initially prepared in the state $\ket{nS,n'P}$ will evolve into the state $\ket{n'P,nS}$ and back: the two atoms exchange energy corresponding to the difference between the energy of the states $\ket{nS}$ and $\ket{n'P}$. This exchange mechanism, also known as "flip-flop", exhibits a $1/R^3$ scaling law and is called dipolar interactions.

When pair states of same energy are not directly coupled to each other by $V_{dd}$, they can interact via second-order processes. A common example is the coupling of the pair state $\ket{nS,nS}$. It is dipole-coupled to states with opposite parity, such as $\ket{n'P,n''P}$, which are separated in energy by $\Delta(n',n'')$. At the second order of perturbation theory, each of those states induces an energy shift proportional to $|\bra{nS,nS}V_{dd}\ket{n'P,n''P}|^2/\Delta(n',n'')$~\cite{reinhard2007level}. In total, the pair state $\ket{nS,nS}$ experiences a van der Waals energy shift $ U_{vdW} = C_6 (n)/R^6 $, where $C_6 (n)$ is a prefactor that increases dramatically with $n$, the principal quantum number~\cite{beguin2013direct}.
\end{shaded}

Another example of spin models that can be realized is the XY Hamiltonian
\begin{equation}
\mathcal{H}(t)= \frac{\hbar}{2} \Omega(t) \sum_j \sigma_j^x -\frac{\hbar}{2} \delta(t)\sum_j \sigma_j^z + 2\sum_{i \neq j}\frac{C_3}{r_{ij}^3} \left(\sigma_i^x \sigma_j^x + \sigma_i^y \sigma_j^y \right).
\label{eq:XY_hamiltonian}
\end{equation}
It naturally emerges when the spin states $\ket{\downarrow}$ and $\ket{\uparrow}$ are two Rydberg states that are dipole-coupled, such as $\ket{nS}$ and $\ket{nP}$~\cite{barredo2015coherent,orioli2018relaxation}. Since the transitions are typically in the tens of GHz range, the effective magnetic field (the two first terms in equation~\eqref{eq:XY_hamiltonian}) are induced by a microwave field instead of an optical laser field. The third term in equation~\eqref{eq:XY_hamiltonian} corresponds to a coherent exchange between the spin states, transforming the pair state $\ket{\downarrow \uparrow}$ into $\ket{\uparrow \downarrow}$. The dipolar interaction that leads to the third term scales as the inverse to the distance $r_{ij}$ to the power of 3 (see Box 2). The exchange interaction is very well suited to study frustrated quantum magnets~\cite{balents2010spin} or to investigate excitation transport~\cite{gunter2013observing}, especially in the context of photosynthesis to understand how the light energy is carried towards the reaction center in light-harvesting complexes~\cite{clegg2006history}. Associated with controllable geometries of the qubits in the register, it can also tackle problems of conductivity in topological materials like organic polymers~\cite{deLeseleuc2019observation}.\\

Combining various pairs of states and playing with the geometry of the spins in the register, neutral atom QPU allows to implement a broad variety of spin Hamiltonians, such as the XXZ model~\cite{bettelli2013exciton,signoles2019observation} or Hamiltonians with non-spin conserving terms~\cite{whitlock2017simulating}. As a very important tool for analog quantum processing, our device also gives the possibility to apply simultaneously global and local addressing beams. It can be used to introduce controllable disorder, which is known to dramatically influence the behaviour of many-body systems~\cite{yao2014many,nandkishore2015many,turner2018weak,2016Gogolin}.\\

Among all the candidates for analog quantum processing, Rydberg atoms in optical arrays are particularly well suited because they offer a very favorable quality factor $\mathcal{Q} \sim 10^2$, given by the ratio of coherent evolution rate over the decoherence rate. Indeed due to their huge electric dipoles, Rydberg interaction energies are typically on the tens of MHz range, at least two orders of magnitude larger than the energy scales associated to decoherence. This gives an estimate of the number of quantum operations that can be realized before the system loses its quantum properties and becomes classical. This quality factor can even be further increased, for instance by trapping the Rydberg atoms to reject the dephasing introduced by their Doppler effect~\cite{barredo2020three}.

\subsubsection{Prospects for improving the QPUs performances\label{sec:ImprovedPerformances}}

\begin{figure}[h!]
\center
\includegraphics[scale=0.45]{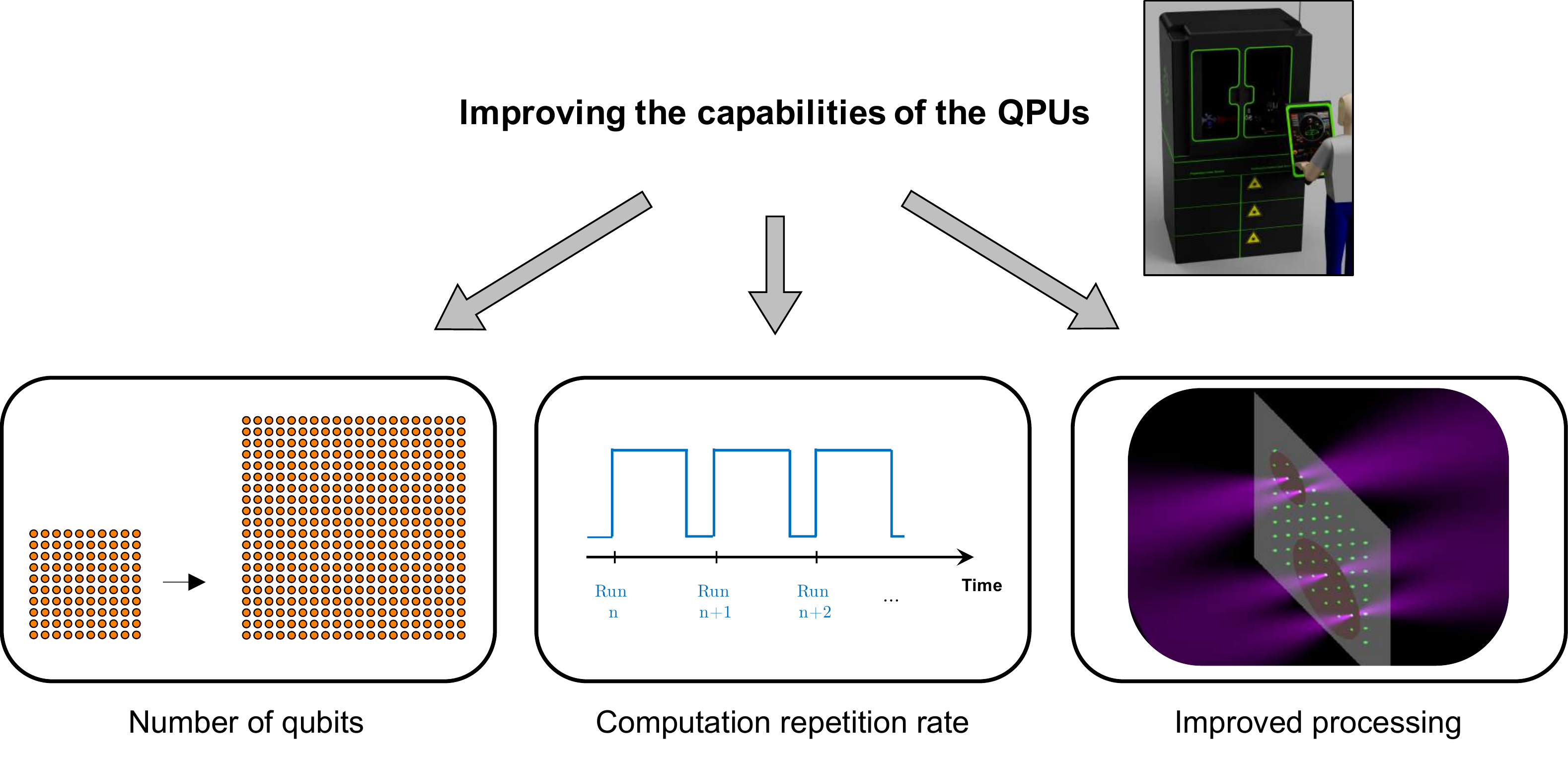}
\caption{The three main axes of hardware developments to improve the performances of the QPUs.}
\label{fig:HardwareDev}
\end{figure}

The capabilities of a quantum processor are directly related to a number of parameters such as the number of available qubits, the repetition rate of the computation, the qubit connectivity in the register or the fidelities of elementary operations, as illustrated in Fig.~\ref{fig:HardwareDev}. An important goal is to identify, study thoroughly and implement technologies to improve on each of those features beyond current performances. We present here a brief overview of the prospects in terms of hardware development.


First, the number of qubits available for computation is mostly limited by the current trapping laser system. On the one hand, the development of new laser systems delivering much higher optical power would enable the generation of many more optical tweezers. Together with an improved imaging system based on state-of-the-art microscope objectives, this would allow to scale the size of the register up to a few thousands of qubits. At this point, the lifetime of the atoms forming a sub-register configuration also becomes a limitation. Because the latter is set by the residual pressure in the vacuum chamber, it motivates another major future development: the design of QPUs in compact cryogenic environments. \\

The other way to overcome the limitation set by the residual pressure is to reduce the duration of each operation occurring in a computation cycle with respect to the atoms lifetime (see Sec.~\ref{sec:register}). First, the development of high-flux atomic sources would allow to reduce the time required to load the registers. Then, a set of new hardware and techniques would be combined to reduce the time required to assemble a defect-free sub-register of atoms, and the time to acquire images. Overall, the repetition rate is expected to improve by an order of magnitude. Even though the current acquisition capabilities enable the realization of most experiments, the overall increase of the repetition rate will facilitate the implementation of procedures that require a very large number of repetitions, such as variational algorithms.\\

Finally, important efforts need to be focused on improving the manipulation of the qubits and reducing the error rate of the quantum operations. In the next generation of neutral atom processors, the electrostatic environment of the atoms will have to be designed with special care to reduce even further the decoherence induced by parasitic charges in the atoms surroundings. On the other hand, because the optical pulses required for computing are generated by radio-frequency pulses, the development of high-bandwidth low-noise electronics would improve the quality of the QPU operations. In particular, agile arbitrary waveform generators (AWGs) with high resolution and large sampling rate would allow to tailor the pulses parameters for optimal control which will boost even further the gate fidelities.\\

Beyond lowering the error rate of successive operations, another way to reduce the effect of decoherence is to shorten the overall computation time. On the one hand, our neutral atom device is well suited to develop tools allowing to perform many operations at the same time (parallelization). On the other hand, the number of operations for a given algorithm can also be reduced with a larger connectivity between the qubits. Several techniques based on pulsing either static or microwave electric fields can be developed to extend the connectivity in the register.


\section{Applications of a neutral atom QPU\label{sec:applications}}

The first natural application of a fully programmable neutral atom Quantum Processing Unit (QPU) is to explore and solve complex quantum phenomena across many areas of science, from the behavior of solid-state materials to chemical and biochemical reaction dynamics. By piloting quantum entanglement and superposition, one reproduces in the device the key elements that are thought to be sufficient to explain such physical phenomena. In that sense, the quantum device acts as a \textit{simulator} of fundamental natural processes and can be used to foster scientific discovery, at a very reduced computational cost. Due to their quantum nature, the scientific problems explored in this Quantum Simulation framework are not solvable easily on classical devices. At the origin of the difficulty lies the exponential scaling of the size of the Hilbert space with the number of interacting particles. 
In this regard, a quantum advantage has already been reached in the study of quantum magnetism\,\cite{Bernien17,Sylvain_PRL_2018}. \\

Beyond the simulation of scientific processes, neutral atom processors can be used to solve hard computational problems, for which classical computers are inefficient. Most of the quantum algorithms tackling computational problems, such as Shor's algorithm, have been designed to be executed on an ideal quantum computer without errors. To reach such a perfect device in practice, one needs to correct the errors that naturally occur at the hardware level with dedicated error correction procedures. By partially measuring some parts of the system, these procedures can detect if errors have occurred in the course of the computation and actively correct them on the fly. Well-known examples include the surface code\,\cite{Bravyi1998,Dennis02,Fowler12}. Architectures that implement error-correction procedures are called fault-tolerant; and they will not be available in the near-future, due to the large overhead in the number of qubits necessary to implement error correction. Even though fault-tolerant quantum computing is still many years away, recent results from Google\,\cite{Arute2019} have shown that devices without error-correction and with a relatively modest number of qubits can already outperform any classical supercomputer for specific computing tasks. Applications in this Quantum Computing framework notably include finding approximate solutions to hard combinatorial optimization problems, solving non-linear partial differential equations or enhancing the performances of Machine Learning procedures.\\

While the Quantum Simulation and Computing frameworks which will be presented respectively in \ref{sec:qs} and \ref{sec:qc} are conceptually different, it is important to realize that they are implemented on the exact \textit{same} physical platform. 

\subsection{Quantum Simulation\label{sec:qs}}

Quantum Simulation consists in studying a synthetic quantum system with the QPU, by implementing a model of interest for which no other way to solve it is known. The model may be an approximate description of a real material or molecule, but it can also be a purely abstract one. There are several approaches to Quantum Simulation, which we describe below. 

\subsubsection{Quantum Simulation on neutral atom devices}

As described in Section \ref{sec:arrays}, neutral atom devices can natively realize quantum spin Hamiltonians (see Eqs. (\ref{eq:ising_hamiltonian}) and (\ref{eq:XY_hamiltonian})). This situation, where one sets the Hamiltonian to exactly mimic the problem under consideration is called Analog Quantum Simulation. This analog setting is not fully general, but corresponds to relatively low requirements in terms of control and operation count, making it the most competitive approach to date. In addition, continuous drive of the system by an external source can extend the variety of phenomena that can be simulated through Floquet engineering\,\cite{Takashi19}. Even though neutral atom devices can reproduce many models of interest, the range of applications of this approach remains limited. To overcome this limitation, another framework has recently emerged\,\cite{kokail_self-verifying_2019}, consisting in using the QPU in tandem with a classical computer. In this hybrid approach, a classical computer is used to optimize a relevant cost function related to a target Hamiltonian, from measurements of highly entangled states prepared on the QPU. In this procedure, the target Hamiltonian never needs to be physically realized with the device. More details about these hybrid classical-quantum procedures are given in Box 3. \\

While hybrid methods greatly extend the simulation capabilities, it is believed that the only target Hamiltonians that can be efficiently simulated are the ones which share the same symmetries as the resource Hamiltonians that can be implemented in the quantum register\,\cite{kokail_self-verifying_2019}. Some models are then intrinsically hard to simulate with neutral atom devices. For those models, the solution consists in using quantum circuits made of gates\,\cite{Lloyd96}. This final approach, called Digital Quantum Simulation, is known to be universal, in the sense that any physical model of interest can be studied. \\


\begin{center}
\begin{tabular}{  |m{4cm}| m{5cm} | m{5cm} |}
\hline
& \textbf{Analog Quantum Simulation} & \textbf{Digital Quantum Simulation}  \\ 
\hline
\textbf{Resource used for simulation} & Hamiltonians    & Gates\\
\hline
\textbf{Key advantages} & Promising hybrid quantum-classical approaches    & Universal approach\\ 
\hline
\textbf{Shortcomings} & Limited number of available configurations    & Requires a large number of gates\\
\hline
\textbf{Status} & Quantum advantage already achieved    & Academic research\\ 
\hline
\end{tabular}
\end{center}


All of these approaches can be implemented on neutral atom QPUs, in various fields of applications, as we describe below.

\definecolor{shadecolor}{rgb}{0.78,0.78,0.78}
\begin{shaded}
\noindent{\bf Box 3 $|$ Variational procedures with Quantum Processing Units}
\end{shaded}
\vspace{-9mm}
\definecolor{shadecolor}{rgb}{0.835294,0.835294,0.835294}
\begin{shaded}
The general aim of these methods is to optimize for a given objective function, and the key idea is to use both a quantum and a classical processor in conjunction, with only a minimal number of operations realized on the quantum processor. In this hybrid framework, the sole use of the quantum processor is to prepare a given quantum state, parameterized by a set of variational parameters. One then extracts from this state an estimate of the objective function through repeated measurements. This estimate is in turn used as the input into a classical optimization procedure, which returns new variational parameters to prepare the quantum state for the next iteration. This loop is repeated multiple times until some stopping criterion is fulfilled, or the optimization procedure converges. Then, a final estimate of the objective function is output.\\

In this framework, the quantum processor can be seen as a trainable quantum algorithm which output is adapted variationally to best match the task under consideration. Importantly, the variational nature of the procedure renders these algorithms relatively resilient to errors\,\cite{Henriet20}, making them prime candidates for an implementation on NISQ devices. The general principle of these hybrid quantum-classical algorithms is illustrated in Fig. B3.1.

\begin{center}
\includegraphics[width=14cm]{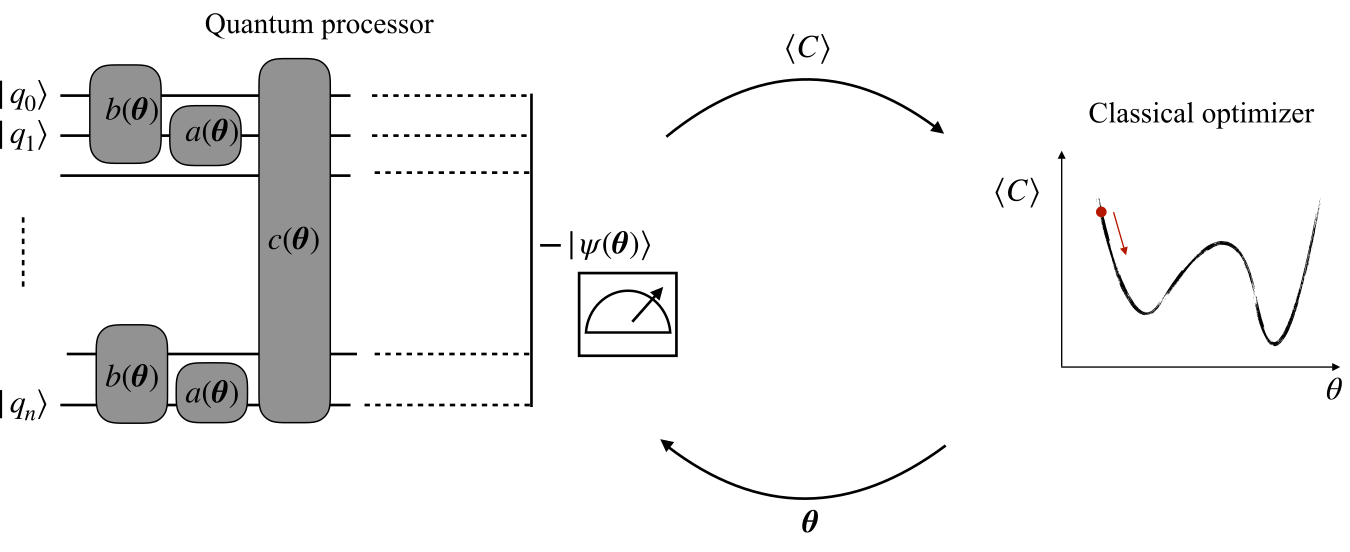}
\end{center}
\begin{small}
\noindent{\bf Figure B3.1 $|$ Principle of hybrid quantum-classical learning algorithms.} 
These algorithms are composed of both a quantum and a classical processor that exchange information within a feedback loop. The quantum processor is used to prepare and measure a $n$-qubit parameterized quantum state. The outcome of the measurement is then used as the objective function in a standard classical optimization procedure, that updates the parameter for the next iteration. The operations on the quantum processor can be of various kinds: single-qubit operations, two qubit operations, or global operations. These operations can be indifferently expressed as digital quantum gates, or Hamiltonian time-evolutions. 
\end{small}\\

This approach might lead to speedups for certain tasks thanks to the “exotic” correlations that can be encoded in the quantum state. Due to its hybrid nature, this technique enhances the potential of the quantum processors available today by drastically reducing the coherence time requirements for the computation. The suitability of such procedures becomes evident when it comes to learning a complex quantum system, such as a molecule’s energy levels in chemistry. 
\end{shaded}

\subsubsection{Applications of Quantum Simulation}

Quantum Simulation applications with arrays of neutral atoms include all of many-body physics, the field that studies the behavior of ensembles of interacting quantum particles. This is a very broad area encompassing almost all condensed matter physics and quantum chemistry\,\cite{Bauer2020} but also nuclear and high-energy physics. 

\paragraph{Condensed-matter physics --}
By allowing the simulation of quantum spin systems, neutral atom devices will open up a variety of new opportunities in Condensed-matter physics. Spin models have been extensively studied in the last 60 years in various contexts, such as magnetism and excitation transport. However, many important open questions remain the subject of active research, such as the nature of the phase diagram when the spins are placed in arrays featuring geometrical frustration, the dynamics of the system after a sudden variation of one parameter of the Hamiltonian\,\cite{Eisert2015}, the role of disorder in the couplings, or their combination with situations where topology plays a role. In the past few years, several important published studies\,\cite{Lienhard18,de_Leseleuc19} have shown that neutral atom devices developed at the Institut d'Optique Graduate School (IOGS) can solve some of these scientific questions. \\

Beyond quantum spin systems, arrays of atoms can also bring new insight into other solid-state systems of interest, like electronic systems. There indeed exist explicit mappings between spin models, that can be implemented natively with neutral atom devices, and electronic Hamiltonians, known as the Jordan-Wigner\,\cite{JW,JW-Nielsen} and the Bravyi-Kitaev\,\cite{Bravyi-kitaev} transformations (see Box 4 for details about the Jordan-Wigner transform). Future investigations along those lines will allow for the study of new materials that will potentially offer unprecedented functionalities for energy transport and storage, or exhibit transformative properties such as high-temperature superconductivity.\\ 

\begin{figure}[h!]
\begin{center}
\includegraphics[width=7cm]{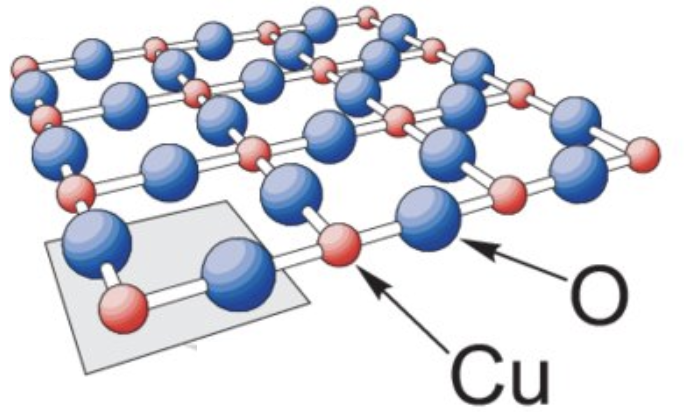}   
\includegraphics[width=6cm]{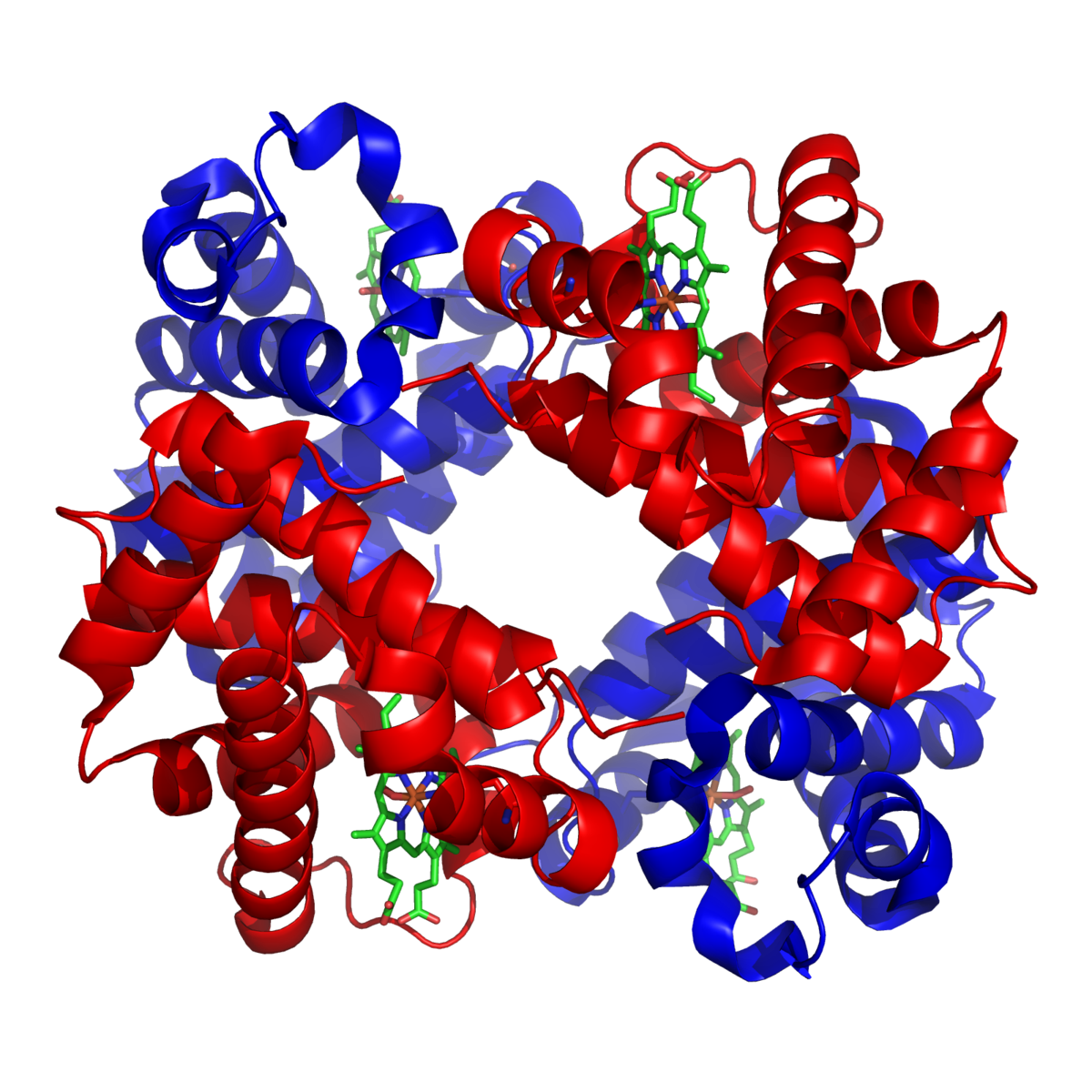}  
\end{center}
\caption{Applications of neutral atom devices to study quantum matter and biochemistry. The left panel shows the copper-oxide plane of a high temperature superconductor\,\cite{Mishonov02}. One models such systems by many-body electronic Hamiltonians (Fermi-Hubbard), that can be mapped on spin systems through well-known transformations. For this application, numerical simulations suggest that a NISQ device able to preserve its coherence for circuit depths of a few hundreds could achieve a quantum advantage over the best known classical methods\,\cite{Clade19}. 
The depth of a circuit corresponds to its total duration, expressed in units of a single gate duration and assuming the ability to carry out operations in parallel. The implementation of parallel gates associated with a relatively modest increase of our gates' fidelities will allow neutral atom devices to go beyond this threshold. The right panel shows the structure of Hemoglobin. Quantum Simulation of the active site of the protein can bring new insight on its reactivity.   }	
\end{figure}

\paragraph{Quantum chemistry --} The ability to simulate electronic systems extends to quantum chemistry and biochemistry problems. While classical computational mechanics is adequate for describing most of the properties of these systems (e.g. with molecular dynamics), the incorporation of quantum effects is instrumental in understanding some physical processes at the microscopic level\,\cite{Melcr19}. The incorporation of many-body quantum effects then allows us to refine models and better understand the reactivity of some molecules, by providing a more complete modelling of the electronic degrees of freedom of the molecules' active sites. This kind of study generally amounts to characterizing the low lying eigenstates of a very large electronic Hamiltonian. Quantum approaches to find eigenvalues have previously relied on the quantum phase
estimation (QPE) algorithm\,\cite{Nielsen}, which offers
an exponential speedup over classical methods,but remains unpractical for NISQ quantum devices without error-correction. An alternative approach was proposed, leveraging the capabilities of quantum hardware through a variational procedure\,\cite{Peruzzo14,McClean16}. 


\definecolor{shadecolor}{rgb}{0.78,0.78,0.78}
\begin{shaded}
\noindent{\bf Box 4 $|$ Jordan-Wigner transform}
\end{shaded}
\vspace{-9mm}
\definecolor{shadecolor}{rgb}{0.835294,0.835294,0.835294}
\begin{shaded}
We illustrate here how to map 1D chain of electrons (fermionic particles) onto a 1D chain of qubits (spin-1/2 particles described by standard Pauli operators, \textit{i.e.} hard-core bosonic particles).\\

Computing the anti-commutator of the two spin operators $\sigma^+_j$ and $\sigma^-_j$ at the same spatial site $j$, we find $\{\sigma^+_j,\sigma^-_j\}=1$, similarly to electronic creation and annihilation operators. However, for different sites, spin operators commuter $[\sigma^+_j,\sigma^-_i]=0$ in contrast to electron operators which anti-commute.\\

Jordan and Wigner introduced in Ref.\,\cite{JW} a transformation that recovers the true fermionic commutation relations from spin-1/2 operators. More specifically, this transformation reads 
\begin{align*}
    &\sigma_{j}^{\dagger }=e^{\left(-i\pi \sum _{k=1}^{j-1}f_{k}^{\dagger }f_{k}\right)}\cdot f_{j}^{\dagger }\\
 &\sigma_{j}=e^{\left(+i\pi \sum _{k=1}^{j-1}f_{k}^{\dagger }f_{k}\right)}\cdot f_{j}\\
  &\sigma_{j}^{\dagger }\sigma_{j}=f_{j}^{\dagger }f_{j}.
\end{align*}
The additional factor above, also named the Jordan-Wigner string, introduces a phase factor at each site, which depends on the density of excited particles located on its left. While we presented its 1D-version, extensions to this transformation also exist in higher dimensions. \\

The mapping above can be used to map any electronic Hamiltonian from condensed-matter physics or quantum chemistry onto a spin (qubit) Hamiltonian. This is the first step of all Quantum Simulation procedures of electronic systems with neutral atom processors. In 2D, this mapping comes with an extra difficulty as simple local
fermionic Hamiltonians are mapped to non-local spin Hamiltonians and vice versa.
\end{shaded}

\paragraph{High-energy physics, nuclear physics and cosmology --} The great control over individual particles allowed by modern quantum simulators enables to realize lattice gauge theory models in practice. Gauge theories are important in particle physics, and include the prevailing theories of elementary particles: quantum electrodynamics, quantum chromodynamics and the standard model of particle physics. Lattice gauge theory is the study of gauge theories on a spacetime that has been discretized into a lattice, which applications also extend to Condensed-matter and nuclear physics\,\cite{Cloet19}. In that respect, one pioneering experiment studied the Lattice Schwinger model with trapped ions \cite{kokail_self-verifying_2019} using a variational approach. Such studies could be implemented on neutral atom devices, which also provide the ability to explore the behavior of 2D lattice models, as noted for example in Ref.\,\cite{Celi2019}. As such, neutral atom devices constitute an appealing platform for testing fundamental theories of physics in the high-energy regime, with a development cost four orders of magnitude lower than a particle collider.\\

We have seen above some of the applications of neutral atom quantum devices for quantum simulation. 

\subsection{Quantum Computing tasks in the NISQ era\label{sec:qc}}

Beyond Quantum Simulation problems, variational procedures with parameterized quantum circuits described in Box 3 can also be applied to more general computational tasks, as was first proposed in Ref.\,\cite{Farhi14} in the context of optimization problems.

\subsubsection{Variational algorithms for computing purposes}

Even though no clear proof of a speed-up with respect to the best classical algorithms exists, several key points have been identified to understand and enhance performances, and the route to quantum advantage using variational procedures for computational problems seems within reach.\\

One particularly promising opportunity for improvement is the implementation of hardware-efficient variational algorithms, that maximally exploit the strengths of available quantum computing devices. Indeed, not all quantum hardware is equivalent, as the set of operations that can be natively implemented depends on the chosen qubit technology (superconducting circuits, neutral atoms, ions, photons, …). On a neutral atom platform, one important example is the native resolution of a well-known graph problem, the Maximum Independent Set (MIS) problem (see details in Box 5). This problem, which has various direct applications in network design\,\cite{Hale} or finance\,\cite{Boginski2005}, becomes hard to solve on a classical computer when the size of the graph grows\,\cite{Garey79}. More generally, solving efficiently the MIS problem would provide a way to solve any interval scheduling problem, with applications in many fields (telecommunication, computing tasks allocation in HPC, or even satellite photography to name a few). \\

\definecolor{shadecolor}{rgb}{0.78,0.78,0.78}
\begin{shaded}
\noindent{\bf Box 5 $|$ Hardware-efficient implementation of a variational algorithm using neutral atom devices}
\end{shaded}
\vspace{-9mm}
\definecolor{shadecolor}{rgb}{0.835294,0.835294,0.835294}
\begin{shaded}
Considering an undirected graph composed of a set of vertices connected by unweighted edges, an independent set of this graph is a subset of vertices where no pair is connected by an edge. The objective of the MIS problem is to find the largest of such subsets. \\

The MIS problem on unit disks graphs can be tackled by using an ensemble of interacting cold neutral atoms as a quantum resource, where each atom represents a vertex of the graph under study. As with any quantum system, the dynamics of the atoms are governed by the Schrödinger equation, involving a Hamiltonian (energy functional) depending on the atomic positions, the electronic energy levels and their interactions. Interestingly, the physical interactions encoded in the Hamiltonian constrain the dynamics to only explore independent sets of the graph under study, then leading to an efficient search in the set of possible solutions\,\cite{Pichler18MIS,Henriet20}, as illustrated in Fig. B5.1. This example underlines the prime importance of targeting the trainable quantum network to the specific task at hand, so that the class of trial quantum states that are generated represents good candidates for solving the problem under consideration.\\

\begin{center}
\includegraphics[width=15cm]{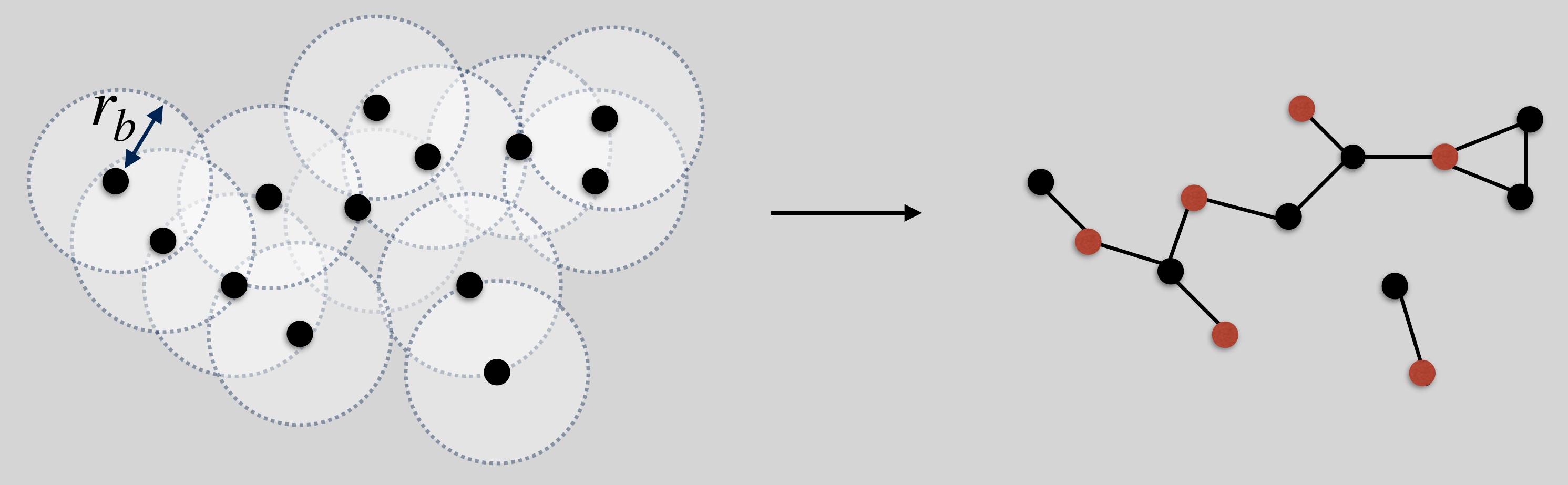}
\end{center}
\begin{small}
\noindent{\bf Figure B5.1 $|$ From a spatial configuration of Rydberg atoms to a unit-disk MIS graph problem.} 
The positions of the atoms, that have two internal energy levels, are chosen to match directly the graph under consideration. The levels of two Rydberg atoms strongly interact if the distance between the atoms is smaller than a typical distance ($r_b$, see left part), resulting in the impossibility for the two atoms to be both in the same state at the same time (phenomenon of Rydberg blockade, as described in Section\,\ref{sec:gates}). This naturally corresponds to the independent set constraint in the graph defined by the atoms, with edges linking atoms that sit at a distance closer than $r_b$. We show on the right the corresponding graph, with the red vertices forming one independent set.
\end{small}\\

As such, well known notions in the field of Quantum Information Processing such as gate fidelity or quantum volume are not always the best criteria for assessing the performances of a given quantum hardware on a specific task. Exploring the same phase space by using only standard digital quantum gates in a nearest-neighbor architecture would be far more demanding than with atoms. A recent preprint from the Google group supported this statement, by finding that the best performance of a variational algorithm for optimization was achieved on a problem that maps directly to the native hardware layout of the device\,\cite{Arute20}. \\
\end{shaded}

By solving the MIS problem for several hundreds sites, neutral atom devices could outperform the best existing classical algorithms for this task\,\cite{Pichler18MIS,Serret20}. Beyond this particular example, progress in the hardware will lead to efficient implementation of variational algorithms on Pneutral atom devices for a variety of other tasks, as we describe below.

\subsubsection{Applications of NISQ Quantum Computing}

Almost all the numerical tasks that are computationally-intensive for standard computing resources are candidates for quantum acceleration using neutral atom devices. This notably encompasses hard optimization problems, the resolution of non-linear partial differential equations or Machine Learning routines.

\paragraph{Combinatorial optimization problems --} The first natural example corresponds to solving NP-hard combinatorial optimization problems, as initially proposed by Farhi et al.\,\cite{Farhi14}. Finding the exact best solution to these problems is not always the main concern, and coming up with an approximate solution in a decent amount of time is often sufficient for application purposes. This fact encouraged the early development of approximate variational algorithms working on noisy devices, as exactness is not strictly required. These algorithms are often referred to in the literature as Quantum Approximate Optimization Algorithms (QAOA).\\

Solving combinatorial optimization problems remains a chief concern in many industrial areas, such as logistics, supply-chain optimization, or network design. In addition, neutral atoms represent a natural architecture for solving optimization problems with multiple-body terms using the Lechner-Hauke-Zoller (LHZ) architecture\,\cite{LHZ15,2017LHZ}.


\paragraph{Non-linear differential equations --} Non-linear differential equations are ubiquitous in all fields of science and engineering. Examples notably include large-scale simulations for reliable weather forecasts, computational fluid dynamics in aeronautics, or even finance. An algorithm was recently proposed to solve non-linear partial differential equations\,\cite{Lubasch20} using a quantum accelerator within a variational procedure. The efficient implementation proposed in that paper makes concrete the prospect of utilizing intermediate-scale quantum processors for solving non-linear problems on grid sizes beyond the scope of conventional computers. 

\paragraph{Machine Learning tasks --} Machine Learning aims at automatically identifying structures and patterns in large data sets. In order to identify these patterns, algorithms often resort to standard linear algebra routines such as matrix inversion or eigenvalue decomposition. For example, support vector machines, one of the most successful traditional machine learning approaches for classification, can be cast to a linear system of equations, and then be solved using matrix inversion. The large dimensionality of the vector spaces involved in these operations makes their implementation at large scale very resource intensive, thus motivating the development of innovative methods to lower their computational cost. Researchers at MIT already proposed by the end of the 90s the use of a quantum processor to perform these linear algebra routines more efficiently (see review \,\cite{WittekBiamonte16}). The discovery of these algorithms triggered a lot of research efforts at the intersection of machine learning and quantum information processing, but the resource requirements for their implementation in term of numbers of qubits and gate fidelities is unfortunately prohibitive for devices that are and will be available in the short term.\\

Motivated by these developments, people tried to find other ways to leverage Machine Learning methods on available NISQ quantum processors, albeit for other purposes than data intensive applications such as the ones mentioned above. New quantum applications for Machine learning are today rapidly developing\,\cite{Schuld2018SupervisedLW,havl19}, and several classifiers have been elaborated\,\cite{Grant18,Mitarai18} as well as sampling methods of classically inaccessible systems. Another promising direction concerns the use of Quantum machine learning for experimental quantum data, opening new prospects in enhanced quantum sensing or metrology.\\

As exemplified in the preceding Section, the high controllability and versatility of neutral atom quantum devices can bring instantaneous value to researchers and practitioners in Quantum Simulation and Quantum Computing in various fields.

\section {Perspectives}
\label{sec:Universal}

The emergence of quantum devices with a few hundred physical qubits (i.e. not error-corrected) opens many exciting perspectives in quantum computing and quantum simulation towards quantum advantage with respect to supercomputers. Among other platforms, fully programmable neutral atom devices display unique characteristics. By controlling quantum entanglement and superposition, either with quantum gates in the digital configuration or with Hamiltonians in the analog configuration, they represent a powerful tool to tackle scientific problems and computing challenges. We explicitly illustrated in the preceding sections how a wide variety of fields will be impacted by the advent of quantum-accelerated computing. We also showed the prospects for future developments at the hardware level, that will allow us to reach the 100 - 1,000 qubits range, and even beyond.\\

In the longer term, neutral atom platforms exhibit many interesting features to push further their computational power and develop new applications. One first appealing direction will be to couple together several processors with optical interconnects. This multi-core architecture would then allow us to substantially scale up the number of qubits available for computation. Not only will the increase of the number of physical qubits allow for the exploration of more industry-relevant usecases, but it will also make possible the implementation of quantum error correcting codes such as the surface code\,\cite{Bravyi1998,Dennis02,Auger2017} on large system sizes. This is one first possible path for the construction of a general-purpose fault-tolerant quantum computing device with neutral atoms.\\

Coupling together distinct processors will require to build coherent interfaces between our atomic qubits and single photons, which in itself constitutes a great experimental challenge. But developing such an efficient interfacing between individual processors and the outer world will also unlock other development avenues.\\

One application of reliable light-matter interfacing consists in using the atomic ensemble as a quantum memory for a photonic qubit. In that framework, quantum information encoded in an incoming photon can be stored in the atomic medium using the phenomenon of electromagnetically-induced transparency (EIT)\,\cite{Harris97,Fleischhauer05}. Under EIT conditions, an atomic ensemble becomes transparent to light and a single photon can propagate inside it without losses under the form of a mixed light-matter excitation called a polariton. The polariton velocity is greatly reduced as compared to the speed of light in vacuum, and can even be temporarily set to zero\,\cite{Bajcsy03}, transforming then the atomic ensemble into a quantum memory. Such a quantum memory could be used for the distribution of quantum information over large distances, which  represents a great challenge due to photon loss. Several so-called quantum repeater protocols\,\cite{DLCZ,Sangouard11} based on atomic ensemble memories have been proposed to overcome this challenge, and experiments are progressing rapidly\,\cite{Maring17,Yu2020}.\\

\begin{figure}[h!]
\center
\includegraphics[scale=0.42]{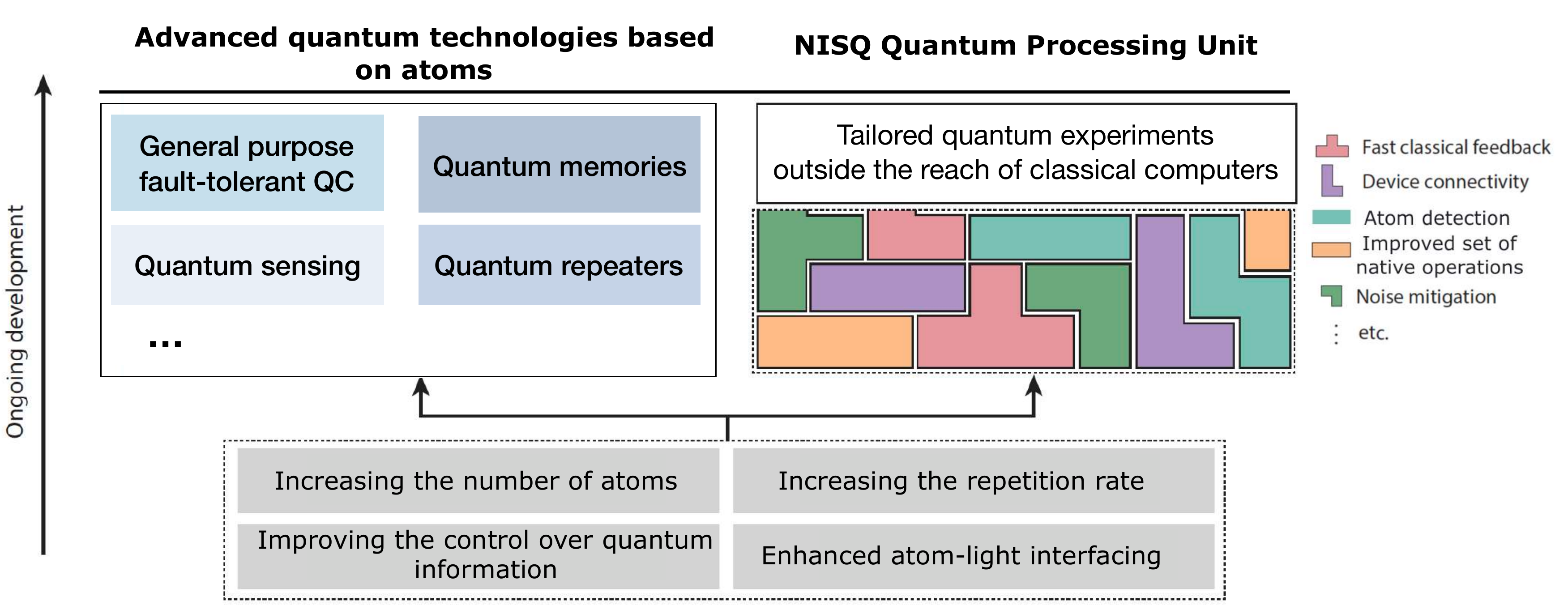}
\caption{Diagram showing the impact of future technological improvements on (i) the quality of our processors in the NISQ era, and (ii) the development of our future atom-based quantum technologies. Diagram adapted from Ref.\,\cite{Kjaergaard2020}.
}
\label{fig:dev}
\end{figure}

In addition, light-matter interfaces could be harnessed to engineer non-linearities between single photons at the single-particle level. This can be done by combining the EIT phenomenon with the strong dipole-dipole interactions in Ryberg media. Under EIT conditions, a single photon can propagate without losses through the atomic medium. When a second photon penetrates inside the medium, the strong dipole-dipole interaction between the atomic components of the two polaritons will result in an effective interaction between the two photons\,\cite{Peyronel2012,firstenberg13}. Such processes constitute a powerful resource for photonic quantum information processing, where quantum logic is applied between photonic quantum bits (see for example Refs.\,\cite{Chang2014,Reiserer2014,paredes14,tiarks16}). In this area of Rydberg non-linear optics\,\cite{Firstenberg_2016,Adams2020}, the Rydberg atoms are no longer the support of the quantum information, but rather act as a source of non-linearity for photonic qubits.\\ 

An interesting recent development about photonic quantum computing is the ability to engineer relatively simple error-correction procedures. The infinite dimensional Hilbert space of photons can be used to redundantly encode quantum information, without having to dramatically increase the number of quantum units. Such realizations have also been proposed with superconducting qubits that are used as resources (ancilla) to perform photonic gates in the microwave frequency range \,\cite{Mirrahimi_2014,Puri2017, alex2020circuit}. The implementation of such photonic (or bosonic) codes would correspond to a second path towards a fault-tolerant architecture.  \\

As briefly illustrated above, and summarized in Fig.\,\ref{fig:dev}, cold neutral atom devices hold great potential for the development of multiple key technologies in the second quantum revolution. 



%
\bibliographystyle{unsrtnat}
\bibliography{refs}


\end{document}